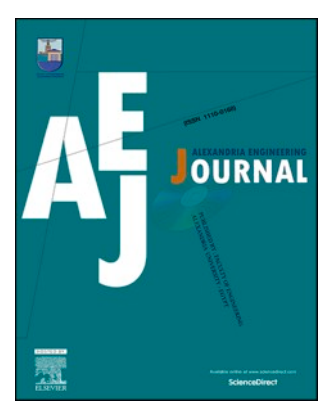

Original article

# Transformer fault diagnosis using an efficient simulation-driven variational quantum classifier with domain-aware feature encoding

Huy Hoang Le, Ba Tu Phung, Dai Huynh, Kim-Anh Nguyen *

*Faculty of Electrical Engineering, The University of Danang - University of Science and Technology, Da Nang 550000, Vietnam*

ARTICLE INFO

*Keywords:*
Simulation-based modeling
Transformer fault diagnosis
Dissolved gas analysis
Variational quantum classifier
Hybrid quantum–classical learning
Quantum circuit simulation

ABSTRACT

Early transformer fault diagnosis is challenged by nonlinear dissolved-gas interactions, overlapping fault signatures, and limited labeled data, while practical deployment further requires reliable performance under realistic computational constraints. This paper presents a simulation-driven modeling framework for dissolved gas analysis-based transformer fault diagnosis, in which a carefully engineered variational quantum classifier (VQC) is employed as the computational core and systematically analyzed through simulation. The framework integrates domain-aware feature modeling derived from Duval geometry with a lightweight two-qubit quantum representation, enabling nonlinear gas-interaction effects to be captured within a shallow parameterized circuit. A hybrid ZX–YY quantum feature map is designed to model non-commuting feature interactions, while a full-entanglement EfficientSU2 ansatz provides adequate expressive capacity under strict resource limits. Model behavior is evaluated using a comprehensive simulation pipeline including noise-aware circuit emulation, cross-dataset validation, and limited hardware-in-the-loop execution, allowing key effects of circuit depth, noise, and optimization strategy to be examined. Simulation results on benchmark dissolved-gas-analysis datasets demonstrate high diagnostic accuracy, strong generalization capability, and robustness to realistic noise levels with minimal quantum resources. The results highlight the effectiveness of simulation-informed modeling for practical transformer diagnostic applications, offering a reproducible and resource-efficient pathway for evaluating quantum-enhanced fault diagnosis methods.

## 1. Introduction

Power transformers are vital assets in modern electrical grids, enabling stable electricity transmission and distribution [1]. However, continuous operation under demanding conditions often exposes transformers to faults such as partial discharge (PD), overheating (T1, T2, T3), and electrical discharges (D1, D2). Undetected faults may escalate into severe failures, costly maintenance operations, and substantial power disruptions, emphasizing the necessity for efficient and accurate diagnostic methods [2,3].

Dissolved Gas Analysis (DGA) has become the primary method for detecting transformer faults by assessing concentrations of gases, hydrogen ($H_2$), methane ($CH_4$), ethane ($C_2H_6$), ethylene ($C_2H_4$), acetylene ($C_2H_2$), and carbon monoxide (CO), in transformer oil [4]. Traditional DGA-based diagnostic frameworks, including IEC 60599, Roger's Ratio Method (RRM), and Duval's Triangle Method (DTM), depend heavily on predefined thresholds and simplistic rules, limiting their adaptability to complex or ambiguous fault scenarios [5].

To overcome these shortcomings, recent advances have integrated Artificial Intelligence (AI) approaches into transformer diagnostics. Techniques such as Fuzzy Logic [6], Support Vector Machines (SVM) [7,8], Decision Trees [9], and Artificial Neural Networks (ANNs) [10,11] have shown improved performance over traditional rule-based methods by capturing intricate nonlinear relationships within DGA data. Nonetheless, classical machine learning models frequently encounter scalability issues due to high computational complexity in large feature spaces and typically require substantial datasets for effective generalization, conditions rarely met in transformer fault diagnostics due to limited labeled data [12].

Quantum Machine Learning (QML), powered by foundational quantum mechanical principles such as superposition, entanglement, and interference, has rapidly evolved into a promising paradigm for enhancing machine learning capabilities beyond classical limits [13–18]. Early advances in this field have demonstrated significant algorithmic benefits: Quantum Support Vector Machines (QSVMs) enable exponential acceleration in kernel evaluations through quantum-enhanced feature maps [19,20], while Quantum Principal Component Analysis (QPCA) achieves quadratic speedups in extracting dominant eigenvectors from large covariance matrices [21,22]. More general quantum neural architectures, including Quantum Neural Networks (QNNs), Parameterized Quantum Circuits (PQCs), and Quantum Generative Adversarial Networks (QGANs), have shown strong potential in modeling nonlinear transformations, capturing high-order feature correlations, and efficiently synthesizing high-dimensional data distributions [23–25].

* Corresponding author.
*E-mail addresses:* lhhoang@powermore.vn (H.H. Le), 105250475@hv.dut.udn.vn (B.T. Phung), 105220381@sv1.dut.udn.vn (D. Huynh), nkanh@dut.udn.vn (K.-A. Nguyen).

These theoretical advantages have motivated an increasing number of industrial applications, particularly in fault diagnosis and condition monitoring. QNN-based frameworks have demonstrated improved parameter efficiency in anomaly detection [26], achieved high accuracy with fewer training iterations in photovoltaic systems [27], and delivered substantial accuracy gains in wind turbine monitoring using hybrid quantum–classical models and QC-CNN architectures [28]. In transformer diagnostics, Variational Quantum Shadow Learning (VQSL) [29] integrates localized variational circuits with classical neural components to enhance scalability and reduce overfitting, particularly under limited or imbalanced DGA data.

Despite these advances, classical machine learning models, and even hybrid quantum-classical approaches, remain constrained when dealing with the intrinsic characteristics of DGA data. Transformer faults often exhibit nonlinear interactions among gas ratios, overlapping class boundaries, and limited sample availability, making fault discrimination inherently challenging. Classical ML relies heavily on linear algebra, convex optimization, and probabilistic modeling; these methods are sensitive to high-dimensional feature correlations and exhibit scalability limitations due to the curse of dimensionality [30]. Hybrid QML approaches mitigate part of this challenge, but they also inherit nontrivial computational overheads, particularly in quantum reinforcement learning (QRL)–based models, where repeated quantum evaluations, large action–state spaces, and iterative policy optimization significantly increase training time and resource consumption. These factors make QRL less practical for real-world transformer monitoring workflows that require rapid, data-efficient learning.

These limitations collectively motivate the use of a more streamlined quantum architecture capable of capturing complex DGA feature interactions without incurring prohibitive computational costs. Variational Quantum Classifiers (VQCs), a prominent QML architecture, encode classical data into exponentially large Hilbert spaces through parameterized quantum circuits, enabling richer geometric structure and more expressive decision boundaries [19,31]. Leveraging entanglement and non-commuting operations, VQCs can efficiently capture intricate feature relationships even with small datasets, an advantageous property for transformer diagnostics where labeled samples are scarce. Recent developments further show that carefully engineered feature maps and ansatz structures can substantially enhance model expressivity while keeping circuit depth shallow, offering a principled, computationally efficient pathway toward improved fault separability within the DGA domain.

These considerations highlight the need for a quantum model that preserves the expressivity advantages of QML while avoiding the substantial computational burden associated with reinforcement-based or deeply layered variational methods. In particular, an effective solution should *(i)* provide strong nonlinear representational capacity for overlapping DGA feature distributions, *(ii)* remain trainable under limited data and shallow circuit depth, and *(iii)* minimize quantum resource requirements to ensure practical deployability on NISQ-era hardware. Addressing these operational requirements motivates an application-driven strategy, systematically tailoring existing VQC techniques to the specific diagnostic needs of power transformers.

Motivated by these challenges, we develop a carefully engineered VQC framework for DGA-based transformer fault diagnosis. The proposed architecture centers on a domain-tailored hybrid quantum feature map informed by Duval geometry, which embeds critical DGA indicators into a two-qubit entangled representation. This design substantially reduces circuit width and depth while preserving the nonlinear expressive power typically associated with larger quantum models. By interleaving non-commuting interactions, the encoder effectively captures the intricate gas-ratio correlations that govern real transformer fault behavior, leading to markedly enhanced class separability in the resulting Hilbert-space embedding.

To further strengthen representational capacity under NISQ constraints, the feature map is coupled with a full-entanglement EfficientSU2 ansatz, optimized via the derivative-free COBYLA algorithm. This combination provides a favorable balance between expressivity and trainability, ensuring stable convergence without incurring the prohibitive computational demands common in QRL-based or deep variational architectures. The resulting VQC yields a lightweight but robust decision model that is well suited for practical deployment.

Comprehensive experiments on benchmark transformer DGA datasets demonstrate that the proposed framework consistently outperforms state-of-the-art classical and hybrid approaches, including SVM variants [32], ANFIS systems [33], deep ANN models [34–37], and recent hybrid ML techniques [38,39]. The classifier attains an exceptional 99.15% accuracy, despite operating with only two qubits and a minimal parameter budget. Beyond raw performance, the model exhibits notable data efficiency, reduced training complexity, and strong scalability potential, making it especially advantageous for real-world monitoring environments characterized by limited, noisy, or imbalanced data.

Overall, this work underscores the practical transformative value of quantum computing for asset-health diagnostics. Rather than claiming fundamental quantum algorithmic novelty, the primary contribution of this study lies in the systematic integration of domain knowledge with practical quantum circuit design. By unifying a principled feature-encoding strategy with a resource-efficient variational architecture, the proposed VQC bridges the gap between quantum theoretical capabilities and operational needs in power-system maintenance, laying the groundwork for next-generation, quantum-enhanced condition-based maintenance solutions.

The rest of the paper is structured as follows: Section 2 reviews fundamental principles of quantum mechanics. Section 3 elaborates on the proposed VQC framework, including feature encoding, quantum circuit architecture, and optimization strategy. Section 4 evaluates experimental results compared to current state-of-the-art methods and the VQSL approach. Section 5 discusses key findings, current limitations, and future enhancements, while Section 6 concludes the paper, suggesting future research directions involving hybrid quantum-classical implementations and deployment on quantum hardware.

## 2. Fundamental principles of quantum mechanics

At the core of quantum computing lies the strange and counterintuitive world of quantum mechanics. Unlike classical bits, which exist in well-defined states of 0 or 1, quantum bits—qubits—exploit the principles of superposition, entanglement, and quantum interference to enable fundamentally new forms of computation [13,14,23]. Three unique properties of qubits are formally described as follows:

a) *Superposition state of the qubit:* In classical computing, a register of $n$ bits can represent one of $2^n$ possible states at any given time. Quantum superposition, however, allows an $n$-qubit quantum system to exist in all these states simultaneously. Mathematically, a qubit $|\psi\rangle$ can be described as:

$$|\psi\rangle = \alpha|0\rangle + \beta|1\rangle, \tag{1}$$

where $|0\rangle$ and $|1\rangle$ are the two possible basis states of the qubit (like 0 and 1 in classical bits). $\alpha$ and $\beta$ are complex numbers representing the probability amplitudes for each state, such that $|\alpha|^2 + |\beta|^2 = 1$. This means the qubit can exist in both $|0\rangle$ and $|1\rangle$states simultaneously, with probabilities given by $|\alpha|^2$ and $|\beta|^2$, respectively.

As a result, quantum computers possess the capability to perform an extensive array of calculations at once, a fundamental advantage that sets them apart from classical computing systems. Fig. 1(a) illustrates a typical quantum superposition state represented on the Bloch sphere, where the qubit occupies a position corresponding to a coherent mixture of classical states.

b) *Quantum Entanglement and Bell states:* Quantum entanglement, one of the most mysterious phenomena in physics, occurs when two or more qubits become intrinsically correlated, regardless of the distance between them. A maximally entangled two-qubit state is known as a Bell state, and one of the most well-known examples is:

$$|\phi^+\rangle = \frac{1}{\sqrt{2}}(|00\rangle + |11\rangle). \tag{2}$$

The state $|\phi^+\rangle$is one of the four Bell states, which together form an orthonormal basis for the Hilbert space $\mathbf{H}_2 \otimes \mathbf{H}_2$ of two-qubit systems. The four Bell states are defined as:

$$\begin{aligned} |\phi^+\rangle &= \frac{1}{\sqrt{2}}(|00\rangle + |11\rangle) \\ |\phi^-\rangle &= \frac{1}{\sqrt{2}}(|00\rangle - |11\rangle) \\ |\psi^+\rangle &= \frac{1}{\sqrt{2}}(|01\rangle + |10\rangle) \\ |\psi^-\rangle &= \frac{1}{\sqrt{2}}(|01\rangle - |10\rangle) \end{aligned} \tag{3}$$

These states describe quantum systems in which the individual qubits do not possess definite states on their own. Instead, the system must be described as a whole. For example, in $|\phi^+\rangle$, if one qubit is measured and found to be in state $|0\rangle$, the other qubit will immediately be found in state $|0\rangle$ as well, and similarly for $|1\rangle$, due to perfect correlation. The factor $\frac{1}{\sqrt{2}}$ ensures that the state is properly normalized so that the total probability is one:

$$\langle\phi^+|\phi^+\rangle = \left(\frac{1}{\sqrt{2}}\right)^2 (\langle 00| + \langle 11|)(|00\rangle + |11\rangle) = \frac{1}{2}(1+1) = 1. \tag{4}$$

This entanglement-based correlation is not merely a curious consequence of quantum mechanics but serves as a critical foundation for quantum information processing. Rather than relying on spatial proximity, entangled qubits maintain a consistent and predictable relationship—an effect famously described by Einstein as “spooky action at a distance”. In quantum computing, this property enables the encoding and manipulation of information across highly correlated quantum states, supporting parallelism and enabling non-classical correlations that outperform classical systems. As illustrated in Fig. 1(b), entangled qubits exhibit robust correlations even when spatially separated, emphasizing the non-local characteristics of quantum mechanics. These correlations are central to a variety of quantum algorithms—such as Grover’s search and Shor’s factoring—and are the basis of key quantum protocols including communication, teleportation, and error correction [40]. By linking qubits in complex, non-local ways, entanglement allows quantum computers to address certain computational problems exponentially faster than their classical counterparts.

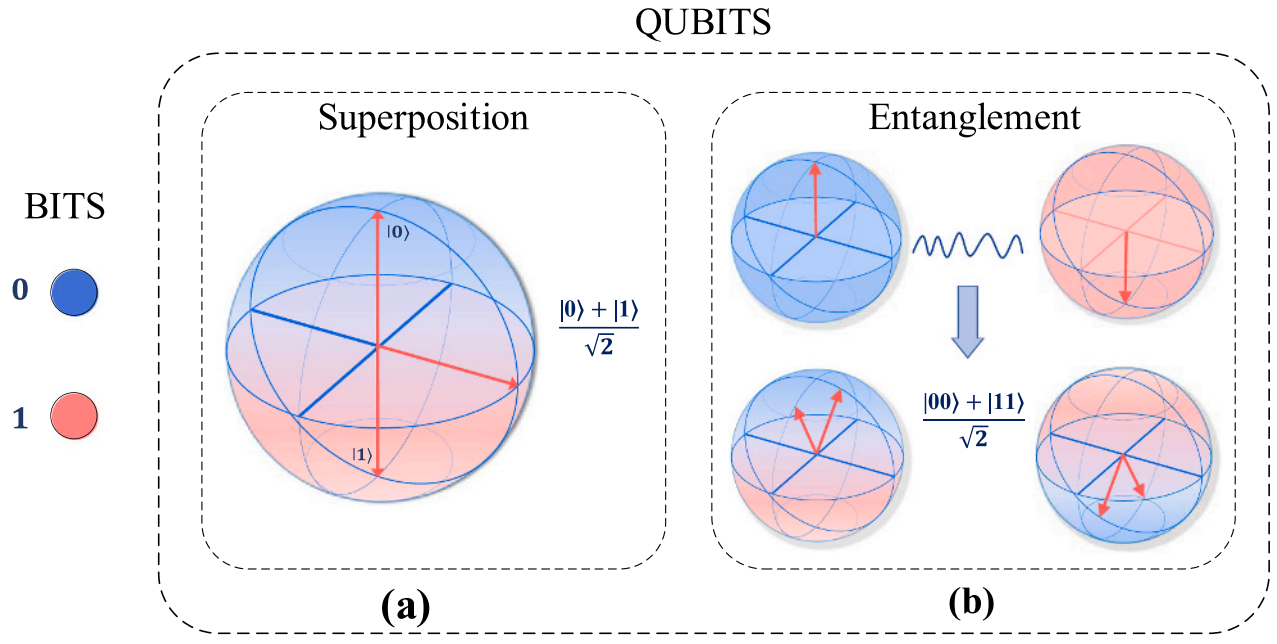

**Fig. 1.** Basic quantum features of qubits.

c) *Quantum interference and computation:* Quantum interference arises from the wave-like nature of quantum states. Probability amplitudes can interfere constructively (reinforcing correct solutions) or destructively (canceling out incorrect solutions). Quantum algorithms, such as Grover’s search algorithm, leverage interference to accelerate problem-solving far beyond classical capabilities.

d) *Quantum measurement and state collapse:* In quantum mechanics, measurement is governed by the Born rule, which states that the probability of observing a particular outcome is given by the square of the amplitude of the quantum state’s projection onto the measurement basis [41]. Given a quantum state represented as a superposition:

$$\Big|\psi\rangle = \sum_{i=0}^{2^n-1} \alpha_i |i\rangle. \tag{5}$$

Where $\alpha_i$ are complex probability amplitudes satisfying the normalization condition:

$$\sum_{i=0}^{2^n-1} |\alpha_i|^2 = 1. \tag{6}$$

Upon measurement in the computational basis $\{|0\rangle, |1\rangle, \ldots, |N\rangle\}$, the state $|\psi\rangle$ collapses to the basis state$|i\rangle$ with probability:

$$P(|i\rangle) = |\langle i|\psi\rangle|^2 = |\alpha_i|^2. \tag{7}$$

This probabilistic collapse is a fundamental property of quantum mechanics, introducing an inherent non-determinism in quantum measurements. Fig. 2 illustrates the quantum measurement process, where an initial quantum state, described by a probability distribution, collapses into a discrete value upon measurement, in which eliminates the superposition of states, forcing the system into a definite state corresponding to the measured value.

## 3. Methodology

This section presents the overall methodology adopted in this study. It is organized into several key components, including feature encoding, ansatz circuit design, VQC model structure, and data preparation. Each component is detailed in the subsequent subsections.

### 3.1. Proposed variational quantum classifier

VQCs belong to the class of hybrid quantum-classical machine learning models, where a quantum circuit is used to encode and process input data, while classical optimization techniques adjust the quantum circuit’s parameters to minimize classification error. The core idea behind VQC is to leverage the high-dimensional Hilbert space of quantum states to represent complex decision boundaries, which may be difficult for classical models to capture efficiently [42]. Fig. 3 illustrates the workflow of a VQC. The process begins with state preparation, where the initial quantum state $|0\rangle^{\otimes n}$is prepared and the input vector $\overrightarrow{x} = [x_0, x_1, \ldots, x_i, \ldots, x_{m-1}]$ is normalized into $\overrightarrow{x}' = [x'_0, x'_1, \ldots, x'_i, \ldots, x'_{m-1}]$, where $i$ is the counting variable from 0 to the total number of input features $m - 1$. Then, a feature map $U(\overrightarrow{x}')$is applied to encode normalized classical input vector $\overrightarrow{x}'$ into a quantum state $\phi(\overrightarrow{x}')$ in a higher-dimensional Hilbert space. The ansatz circuit $U(\overrightarrow{\theta}_k)$ applies parameterized quantum gates (eg., $R_Y(\theta)$, $R_Z(\theta)$), which are optimized to minimize classification error. Here, $k = 1, \ldots, K$ with $K$ is total number of iterations; $j = 1, \ldots, 8 + 4(\text{reps} - 1)$, where reps is total number of ansatz layers; $\vec{\theta}_k = [\theta_0, \theta_1, \ldots, \theta_j]$ is the trainable parameters vector at iteration $k$. After measurement, the output is passed to an optimizer, which updates the $\vec{\theta}_k$ iteratively to refine the decision boundaries. This hybrid approach combines quantum encoding for feature representation with classical optimization techniques to enhance classification performance on complex datasets.

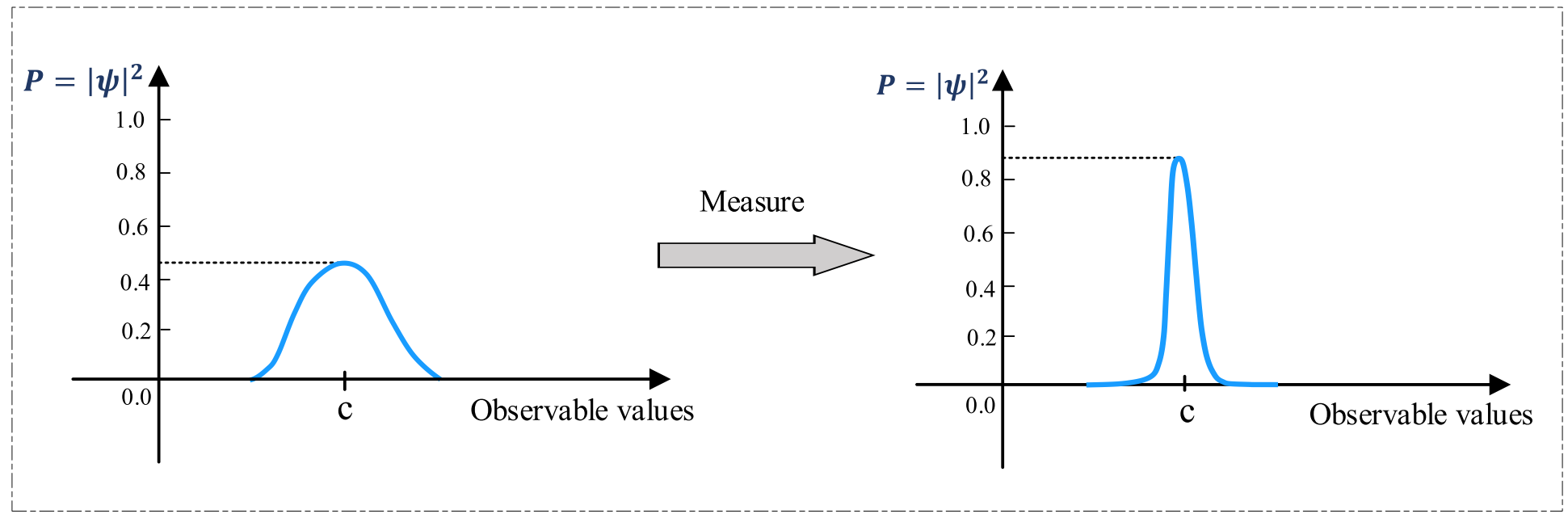

**Fig. 2.** Wavefunction collapsed upon measurement.

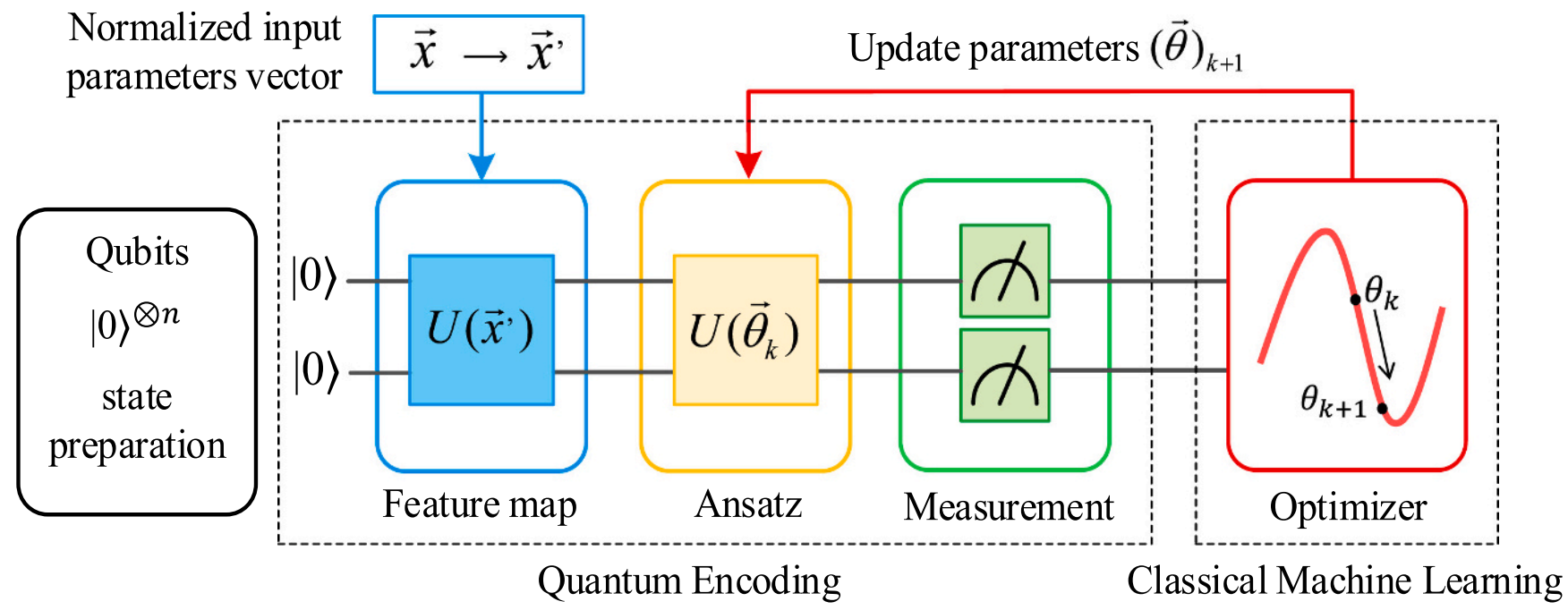

**Fig. 3.** Hybrid quantum-classical training loop in a Variational Quantum Classifier.

#### 3.1.1. Feature map

After the initial state preparation, where qubits are initialized to the basis state $|0\rangle^{\otimes n}$, a feature map is applied to encode the normalized classical input vector $\vec{x}'$. This process transforms the initial state into a quantum state $\phi(\vec{x}')$ in a higher-dimensional Hilbert space [42].

In this study, we implement a quantum feature map designed to encode classical input features into quantum states using two-qubit entangling gates. The circuit utilizes parameterized $R_{YY}(\theta)$and $R_{ZX}(\theta)$, gates which are natively supported in Qiskit but decomposed into hardware-compatible gates such as $R_X$, $R_Z$, Controlled NOT ($CX$), and Hadamard ($H$). The feature map is defined as a 2-qubit system and encodes 4 real-valued features $\vec{x}' = [x'_0, x'_1, x'_2, x'_3]$ through the following structure:

$$|\psi(\vec{x}')\rangle = U_{feature_map}(\vec{x}')|00\rangle = R_{ZX}(x'_3)_{0,1} R_{YY}(x'_1)_{1,0} R_{ZX}(x'_2)_{1,0} R_{YY}(x'_0)_{0,1}|00\rangle, \quad (8)$$

Each gate operates across both qubits, allowing feature interactions and entanglement to be embedded in the quantum state, effectively expanding the expressive capacity of the quantum model. The number "0,1" or "1,0" below the rotation gates is the position of the current qubit acts on other qubit, which means that the rotation gate $R_{YY}(x'_0)_{0,1}$ is currently at qubit 0 and acts on qubit 1, with $x'_0$ is an input parameter. To quantitatively validate how this structural entanglement effectively captures nonlinear feature interactions, detailed mathematical derivations of the closed-form kernel and comprehensive entanglement metrics are provided in Appendix A.

The structure of the proposed entanglement-based quantum feature encoding circuit, which utilizes interleaved $R_{YY}(\theta)$ and $R_{ZX}(\theta)$ interactions to encode four classical features into two qubits, is illustrated in Fig. 4. It is important to note that the visual representation of the quantum circuit may appear reversed relative to the mathematical formulation in Eq. (8), as quantum circuits are conventionally drawn with gate operations applied from left to right, whereas mathematical expressions are typically interpreted from right to left in terms of operator application.

#### 3.1.2. Parameterized ansatz

Following the mapping of classical data to a quantum state, the subsequent procedure involves the application of a PQC, referred to as the ansatz, to transform the input quantum state. The ansatz comprises a series of quantum gates with adjustable parameters, which are optimized during the training process to minimize classification error.

In Table 1, we compare common two-qubit VQC ansatz under a fixed CNOT budget: (i) expressivity (tends to lower $F_2$ with depth), (ii) trainability (gradient variance; distance from 2-designs), (iii) when to use, and (iv) hardware practicality (CNOT count, native compilation).

Following Sim et al. [46] we evaluate ansatz by expressibility/entangling capability and account for trainability limits due to barren plateaus. EfficientSU2 (Ry–Rz with CX) maintains high expressibility at shallow depth without approaching a 2-design, whereas RealAmplitudes prepares real states and plateaus unless extra mixers/re-upload are added. PauliTwoDesign targets a 2-design, maximizing expressibility but increasing barren-plateau risk. These theoretical properties align with our experiments (Table 3 in Subsection 4.2.2), where EfficientSU2 dominates at every depth [43–45].

For this study, the EfficientSU2 ansatz is employed due to its strong entangling capability and shallow circuit depth, making it well suited for NISQ hardware [42]. EfficientSU2, widely supported in Qiskit, consists of layered single-qubit rotations and CX entanglers. Each layer includes:

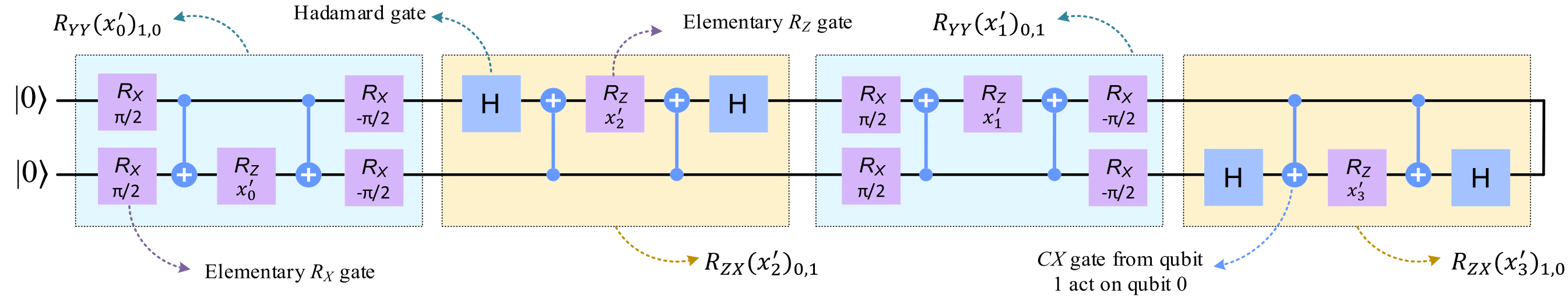

**Fig. 4.** Quantum feature encoding circuit decomposed into native gates, implementing a sequence of $R_{YY}(\theta)$ and $R_{ZX}(\theta)$-inspired interactions for four-dimensional data.

**Table 1**
Ansatz circuit comparison.

| Ansatz | EfficientSU2 | RealAmplitudes | PauliTwoDesign |
|---|---|---|---|
| Params / layer (2q) | ~8 | ~4 | 8–12 |
| 2-qubit gates / layer (approx.) | 1–2 | 1–2 | 1–3 |
| State space | Complex (has phase via $R_z$) | Real (no complex phase) | Complex, near-random ensemble |
| Typical layer structure | $[R_yR_z]^{\otimes 2}$→CNOT→$[R_yR_z]^{\otimes 2}$ | $R_y^{\otimes 2}$ + CNOT | Pauli rotations + random entanglers (targets 2-design) |
| Expressivity & entanglement | Strong non-commuting mixing: expressivity grows smoothly with reps *CX*entangler gives good entanglement [43,44] | Expressivity limited by missing phase; often plateaus with depth unless extra re-upload/ mixing is added [44] | Very high expressivity: ensemble approaches Haar quickly as depth grows. |
| Trainability (barren-plateau risk) | Low–medium at shallow depth (doesn't hit 2-design too fast) [45] | Low (but capacity is limited) | High (prone to barren plateaus at moderate depth) [45] |
| When to use | Small/medium ML tasks needing good accuracy per-*CNOT*; matches encoders that inject phase (e.g., ZX–YY) | Very resource-constrained or very simple data; not ideal when the encoder creates off-basis coherences | Exploring representational limits / randomization; less suitable for small-data DGA (can be over-expressive) |

- Parameterized rotations: Every qubit undergoes$R_Y(\theta)$and $R_Z(\theta)$operations, introducing trainable parameters and expanding the accessible Hilbert-space region.
- Entanglement block: CX gates are applied according to a chosen entanglement pattern (e.g., linear or full), enabling the circuit to capture cross-qubit correlations effectively.
- Post-entanglement rotations: A second set of $R_Y(\theta)$and $R_Z(\theta)$gates further enhances expressivity and supports richer state transformations.

Fig. 5 presents an instance of the EfficientSU2 ansatz, constructed with 2 qubits and 2 repetitions, which highlights the alternating sequence of parameterized single-qubit rotations and entangling *CX*gates. This configuration exemplifies how even a relatively shallow ansatz can induce substantial entanglement and nonlinearity within the quantum state space.

Mathematically, the ansatz applies a unitary transformation $U(\vec{\theta})$ on the feature-encoded quantum state:

$$\left|\psi_{out}\left(\vec{\theta}\right)\right\rangle = U\left(\vec{\theta}\right)|\psi(\vec{x}')\rangle, \tag{9}$$

where $U(\vec{\theta})$ consists of trainable parameters vector $\vec{\theta}$ optimized during training.

### 3.1.3. VQC optimization

a) *Quantum optimization loop in VQC*

Variational Quantum Classifiers rely on PQCs, where a set of trainable parameters $\vec{\theta}$ are embedded into unitary transformations. The goal of optimization is to adjust $\theta_j$ such that the classifier correctly predicts the class labels based on quantum measurement outcomes. The optimization process follows a hybrid quantum-classical loop (as illustrated in Fig. 6), includes three following steps:

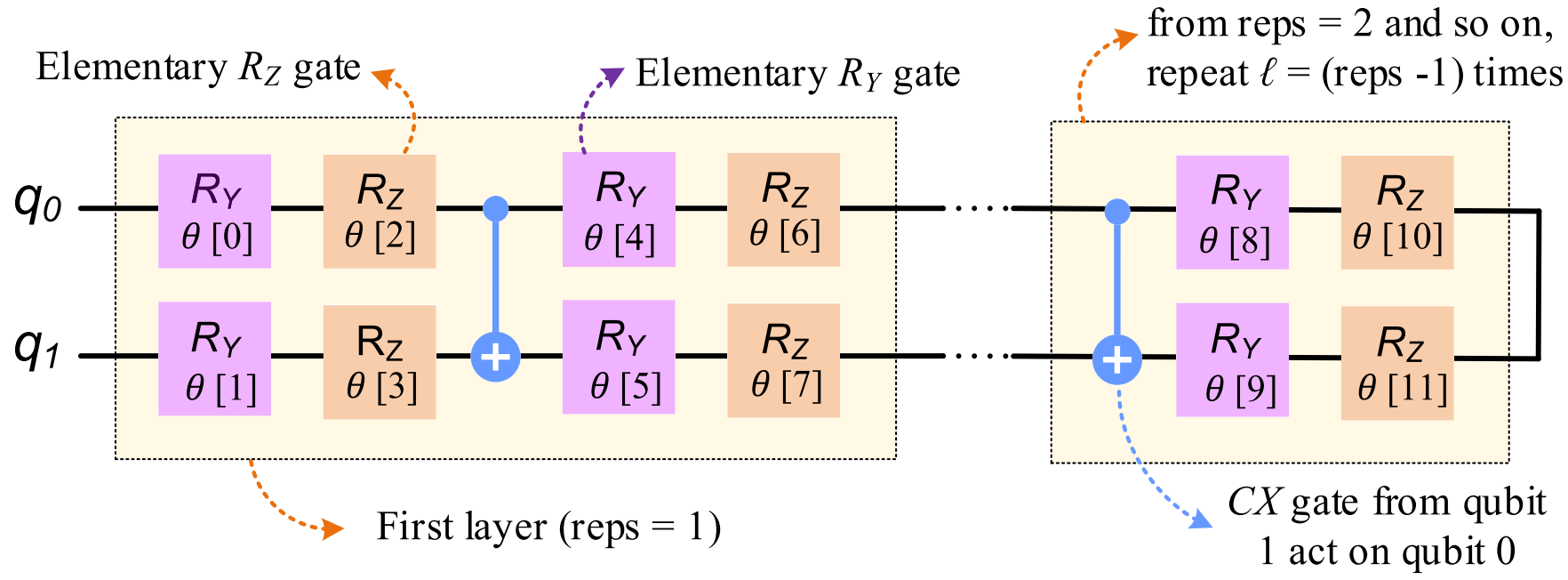

**Fig. 5.** Gate-level representation of the EfficientSU2 ansatz circuit on two qubits, utilizing interleaved $R_Y(\theta)$, $R_Z(\theta)$ rotations and entangling *CX* gates.

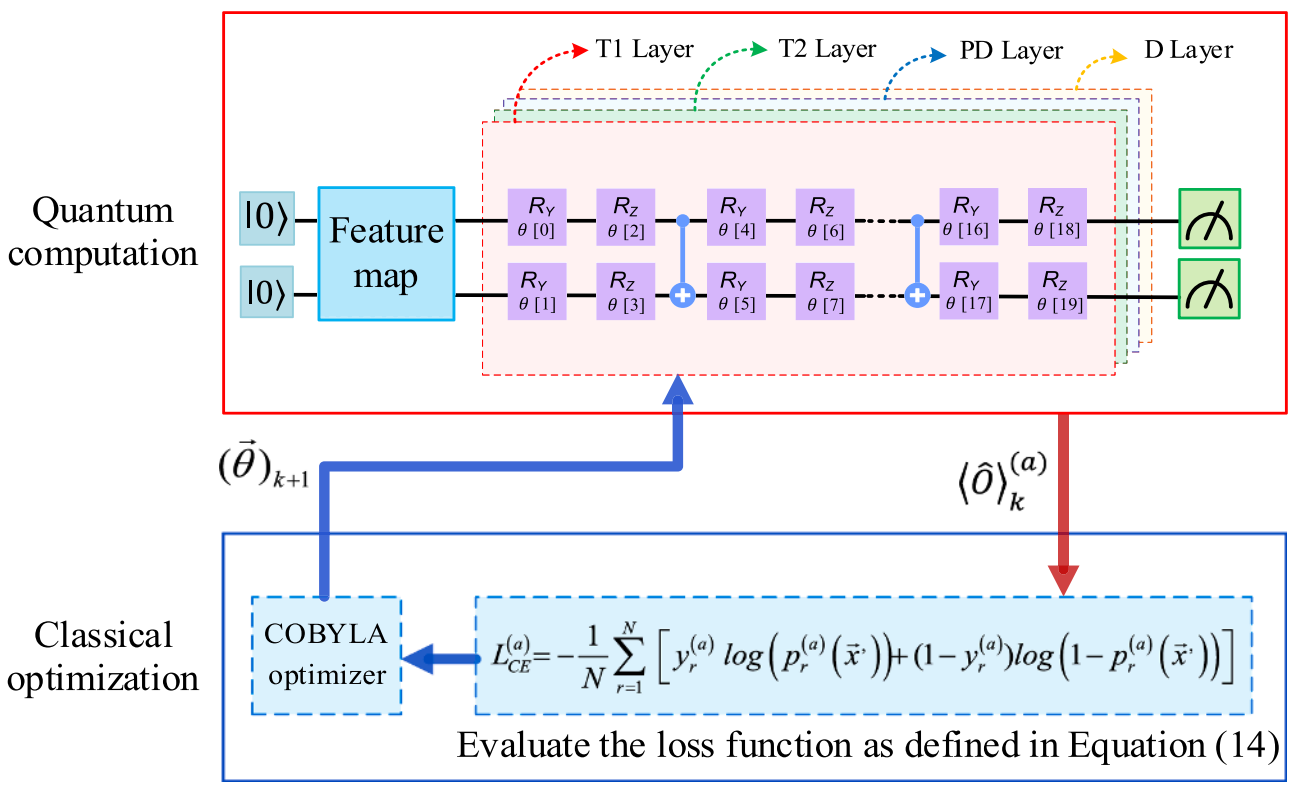

**Fig. 6.** The optimization process employing a hybrid quantum-classical loop.

(i) Step 1 - Quantum Computation: The input classical data is encoded into quantum states using the feature map, and the parameterized quantum circuit transforms the state.
(ii) Step 2 - Measurement: The expectation value of the observable $\hat{O}$ (e.g., Pauli-Z operator) is computed.
(iii) Step 3 - Classical Optimization: A classical optimizer updates the circuit parameters based on the computed loss function. Steps 1–3 are repeated until convergence is reached.

when applying quantum models to supervised learning tasks, objective function can be formulated as a classical loss function over a dataset:

$$C(\vec{\theta}) = \sum_{r=1}^{N} L(y_r, f_{\vec{\theta}}(\vec{x}')_r), \tag{10}$$

where $y_r \in \{0,1\}$ is the ground truth label for the $r-th$sample, $N$ is the total numbers of samples that used to train the model, $f_{\vec{\theta}_j}(\vec{x}')_r$ is the quantum classifier's output for the $r-th$ sample, $L$ is the loss function used for classification.

In this study, we adopt the cross-entropy loss function, which is a widely used objective function for classification tasks, especially in probabilistic and binary classification settings. The cross-entropy loss, denoted $L_{CE}$, is given by:

$$L_{CE} = -\sum_{r=1}^{N} [y_r \ log(p_r) + (1-y_r)log(1-p_r)], \tag{11}$$

where $p_r \in [0,1]$ is the predicted probability of the $r-th$sample being classified as class 1, i.e.,$p_r = f_{\vec{\theta}}(\vec{x}')_r$. This loss function penalizes the deviation of the predicted probability $p_r$ from the true label $y_r$, encouraging the model to output probabilities that are close to the true class distribution.

COBYLA optimizes the loss function iteratively by constructing linear approximations of the objective function and updating parameters accordingly. The iterative process is given by:

$$\vec{\theta}_{k+1} = \vec{\theta}_k + \gamma d_{jk} \text{ with } d_{jk} = \underset{d}{argmin}\ L_{CE}(\theta_{jk} + d) \tag{12}$$

Where, $\vec{\theta}_k$ is the parameters vector at iteration $k$; $\gamma$is the adaptive step size; $d_{jk}$is the search direction of parameter $\theta_j$ at iteration $k$, determined by COBYLA's linear approximation method.

At each iteration, COBYLA adjusts $\vec{\theta}_k$ based on function evaluations, refining the quantum circuit's parameters to reduce classification error. The optimization terminates when the objective function converges or when the predefined number of iterations is reached.

b) *One-vs-Rest VQC classification algorithm*

The One-vs-Rest (OVR) VQC classification framework (Algorithm 1) extends the inherently binary nature of variational quantum classifiers to multi-class problems. Rather than relying on a single multi-class circuit, the method trains one binary VQC per class, each tasked with distinguishing its target class from all others. At inference, every classifier outputs a class-specific score, and the final label is assigned to the class with the maximal decision value. This modular design provides a simple yet effective mechanism for multi-class prediction while preserving the expressive quantum feature encoding and optimization benefits of the underlying VQC architecture.

During the training phase, the algorithm constructs $J$independent binary classifiers $\left\{f^{(a)}_{\vec{\theta}}(\vec{x}')\right\}_{a=0}^{J-1}$, where each classifier $f^{(a)}_{\vec{\theta}}$is trained to distinguish class $a$ from the rest. The label transformation for OVR is expressed as Eq. (25).

The input vector $\vec{x}'$ is encoded into a quantum state $|\psi(\vec{x}')\rangle$ using a custom-designed Feature Map $U(\vec{x}')$, followed by an EfficientSU2 ansatz $U\left(\vec{\theta}\right)^{(a)}$acting on 2 qubits:

$$|\psi^{(a)}_{out}\rangle = \left(U\left(\vec{\theta}\right)^{(a)} \cdot U(\vec{x}')\right)|0\rangle^{\otimes\ 2}. \tag{13}$$

A measurement is then performed on a predefined qubit (e.g., qubit 0), and the output probability $p^{(a)}_r = \langle\psi^{(a)}_{out}|\hat{O}^{(a)}|\psi^{(a)}_{out}\rangle$is used to approximate the likelihood that $\vec{x}'$ belongs to class $a$. The objective is to minimize the binary cross-entropy loss:

$$L^{(a)}_{CE} = -\frac{1}{N}\sum_{r=1}^{N} \left[y^{(a)}_r \ log(p^{(a)}_r(\vec{x}')) + (1-y^{(a)}_r)\ log(1-p^{(a)}_r(\vec{x}'))\right]. \tag{14}$$

Each loss $L^{(a)}_{CE}$is minimized with respect to $\vec{\theta}^{(a)}$using the COBYLA optimizer, with a maximum of 100 iterations per classifier.

For a test sample $\vec{x}'_{test}$, predictions are obtained by evaluating all trained classifiers:

$$p^{(a)}_{test} = f^{(a)}_{\vec{\theta}^{(a)}}(\vec{x}'_{test}), \quad \forall a \in \{0,1,\ldots,J-1\}. \tag{15}$$

A voting-based decision rule is applied:

(i) If any classifier yields $p^{(a)}_r > \tau$ (threshold, typically 0.5), the corresponding class $a$ is selected.
(ii) If multiple classifiers satisfy this condition, the class with the highest probability is chosen:

$$\hat{y} = \underset{a}{\text{argmax}}\{p^{(a)}_r | p^{(a)}_r > \tau\}. \tag{16}$$

(iii) If no classifier yields $p^{(a)}_r > \tau$, fallback to:

$$\hat{y} = \underset{a}{\text{argmax}}\ p^{(a)}_r. \tag{17}$$

### 3.2. Data preparation

The dataset comprises 588 DGA samples drawn from two sources: 470 samples from the dataset in [47] for training (80%) and 118 IEC TC 10 samples [48] for testing (20%). All samples correspond to dissolved-gas measurements in transformer oil. The training set comprises 470 samples exclusively from [47]. The IEC TC 10 dataset (118 samples) serves as a strictly held-out external test set with zero sample overlap. The two datasets differ in provenance, oil-handling practices, and sensor calibration, thereby introducing a realistic domain shift to evaluate true generalization. Furthermore, the StandardScaler is fitted solely on the 470 training samples; its learned parameters $(\mu, \sigma)$ are then applied to transform the 118 test samples, mathematically ensuring no information leakage from the test set to the training set. Table 2 summarizes the distribution of the four fault classes (T1, T2, PD, D), revealing a notable class imbalance—243 D (41.32%), 212 T2 (36.06%), 79 T1 (13.44%), and only 54 PD (9.18%)—equivalent to a ~4.5:1 ratio between the majority (D) and minority (PD) classes. This skewed distribution provides a realistic basis for evaluating the model's robustness to imbalance.

**Table 2**
Distribution of fault types across training and testing datasets.

| Status | Training samples | Percent training | Testing samples | Percent testing | All samples |
|---|---|---|---|---|---|
| T1 | 63 | (13.4%) | 16 | (13.56%) | 79 |
| T2 | 193 | (41.0%) | 19 | (16.10%) | 212 |
| PD | 45 | (9.6%) | 9 | (7.63%) | 54 |
| D | 169 | (36.0%) | 74 | (62.71%) | 243 |
| **Total samples** | **470 (100%)** | | **118 (100%)** | | **588** |
| **Dataset split (%)** | **80%** | | **20%** | | **100%** |

To statistically evaluate the model's robustness and generalization capability, an independent benchmarking approach (cross-site evaluation protocol) was conducted. The model was trained exclusively on the 470 samples, whereas the distinct IEC TC 10 dataset served as an external benchmark for performance testing and comparison with state-of-the-art methods.

Prior to model training and evaluation, the raw dataset, consisting of five key gas ratios, was transformed into reduced feature representations by computing their corresponding coordinate systems. Specifically, the Duval triangle coordinates ($T_x, T_y$) and Duval pentagon coordinates ($P_x$, $P_y$) were derived to facilitate fault classification while reducing the dimensionality of the feature space (as illustrated in Fig. 7). The coordinates were calculated using the following transformations:

Duval Triangle Coordinates [49], where [.] mean gas concentration in ppm:

$$T_x = \left( \frac{[C_2H_4]}{([C_2H_2] + [C_2H_4] + [CH_4]) \times 0.866} + \frac{[CH_4]}{([C_2H_2] + [C_2H_4] + [CH_4]) \times 1.732} \right) \times 0.866, \quad (18)$$

$$T_y = \frac{[CH_4]}{[C_2H_2] + [C_2H_4] + [CH_4]} \times 0.866. \quad (19)$$

In the Duval Pentagon method [50], the coordinates $x_1$ to $x_5$ and $y_1$ to $y_5$ correspond to the five vertices of the Duval Pentagon, each representing one of the key fault gases: $H_2$, $C_2H_6$, $CH_4$, $C_2H_4$, $C_2H_2$. These coordinates are computed by projecting the normalized gas concentrations onto a 2D plane using cosine functions with fixed angles to form a pentagonal shape. This transformation enables the visualization of gas ratios as geometric positions for fault classification. The DPM coordinates are calculated as:

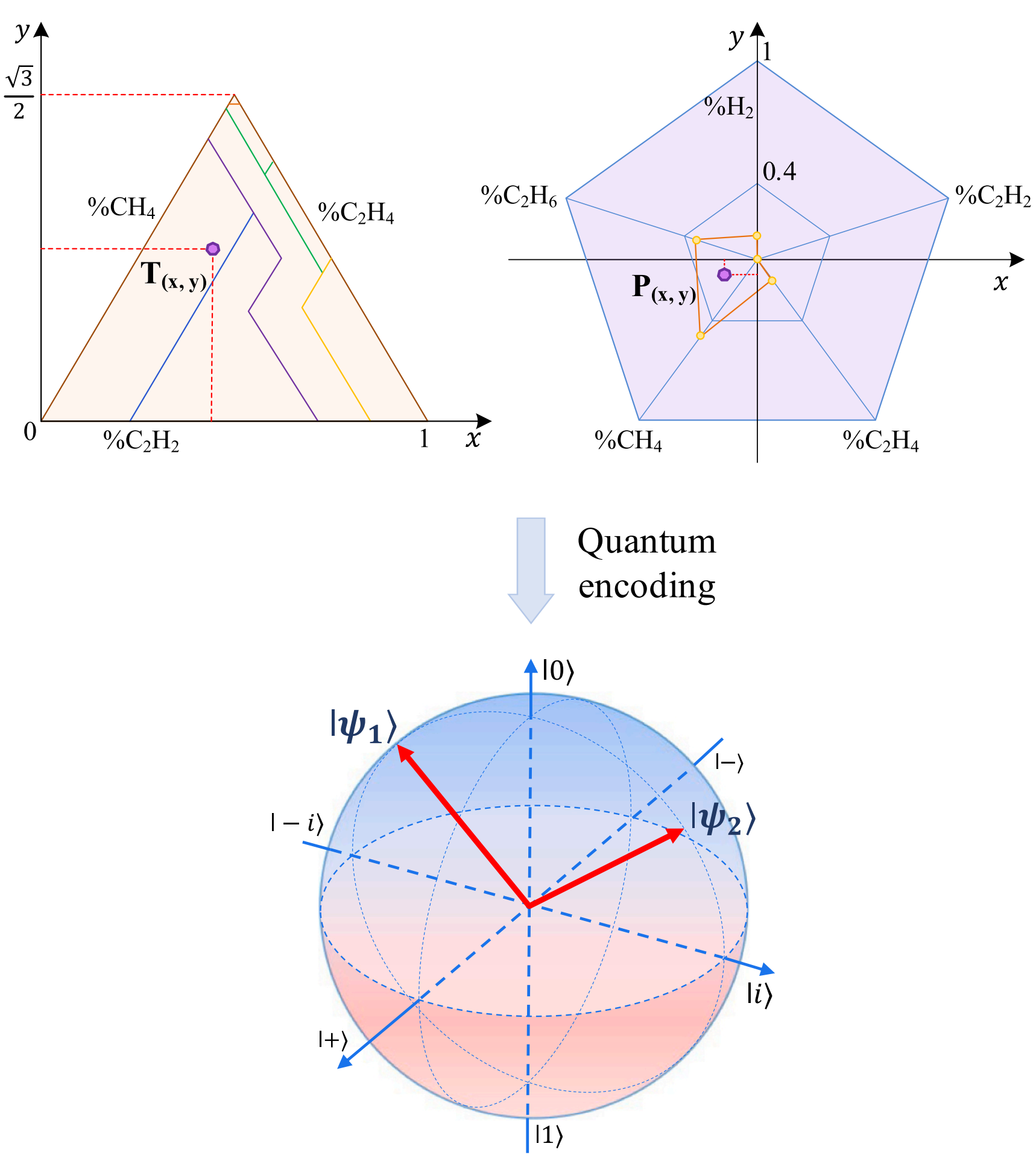

**Fig. 7.** 2D cordinate features tranforms into high-dimensional Hilbert spaces.

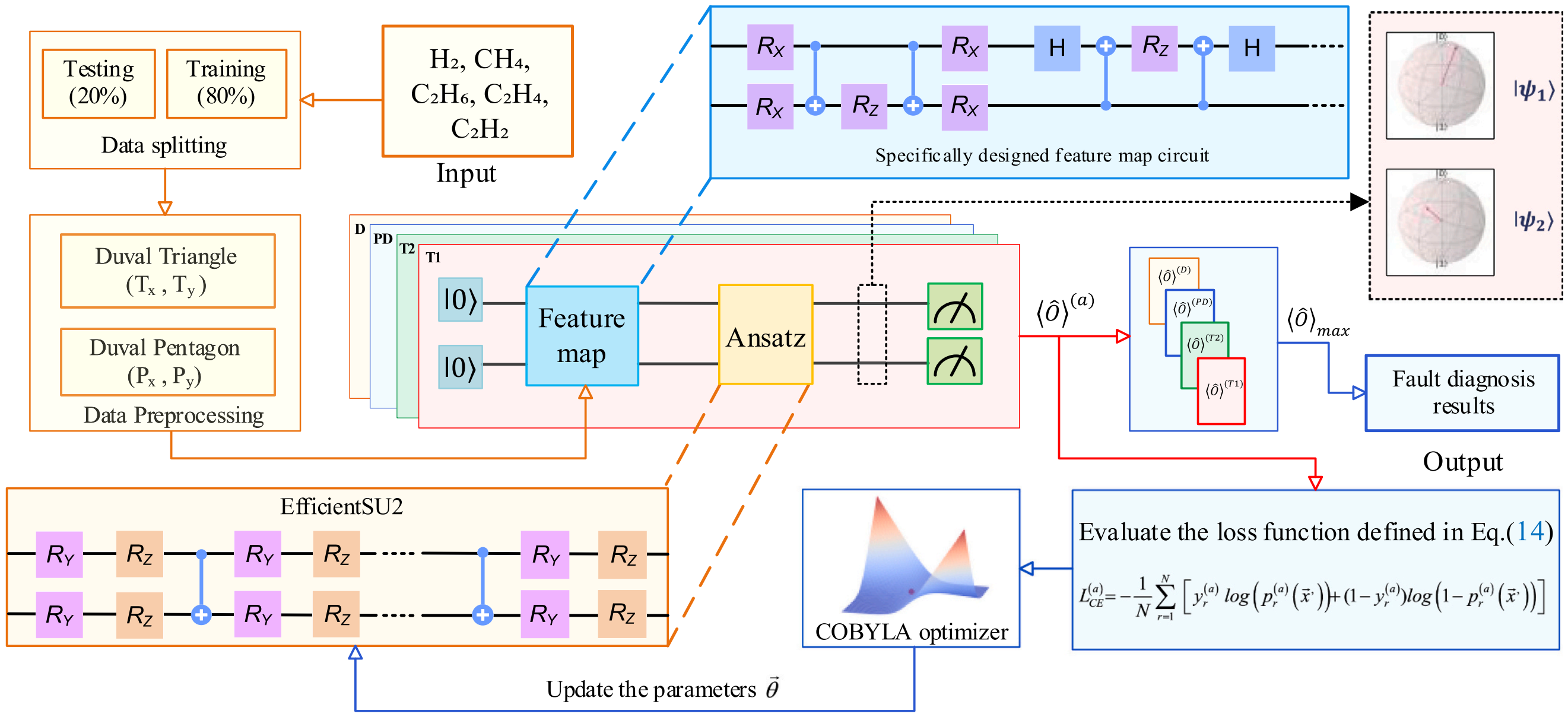

**Fig. 8.** Proposed VQC-based transformer fault diagnosis framework.

$$\begin{cases} x_1 = 0 & y_1 = [H_2] \\ x_2 = [C_2H_6] \times \cos(180^\circ - 18^\circ) & y_2 = [C_2H_6] \times \cos(90^\circ - 18^\circ) \\ x_3 = [CH_4] \times \cos(180^\circ + 54^\circ) & y_3 = [CH_4] \times \cos(90^\circ + 54^\circ) \\ x_4 = [C_2H_4] \times \cos(-54^\circ) & y_4 = [C_2H_4] \times \cos(90^\circ + 54^\circ) \\ x_5 = [C_2H_2] \times \cos(18^\circ) & y_5 = [C_2H_2] \times \cos(90^\circ - 18^\circ) \end{cases}, \tag{20}$$

$$P_x = \frac{\sum_{p=1}^{4}(x_p + x_{p+1})(x_p y_{p+1} - x_{p+1} y_p)}{3\sum_{p=1}^{4}(x_p y_{p+1} - x_{p+1} y_p)}, \tag{21}$$

$$P_y = \frac{\sum_{p=1}^{4}(y_p + y_{p+1})(x_p y_{p+1} - x_{p+1} y_p)}{3\sum_{p=1}^{4}(x_p y_{p+1} - x_{p+1} y_p)}. \tag{22}$$

Following the determination of these coordinates, a normalization step was performed using the Standard Scaler technique to ensure that each feature exhibited a mean of 0 and a standard deviation of 1. Mathematically, each feature in the input parameters vector $\vec{x} = [T_x, T_y, P_x, P_y]$ was transformed using the formula:

$$\vec{x}' = \frac{\vec{x} - \mu}{\sigma}, \tag{23}$$

where $\vec{x}'$ is the normalized input parameters vector,$\mu$ is the mean of the feature, $\sigma$ is the standard deviation of the feature. This process mitigates disparities in scale among the input parameters, thereby enhancing the machine learning performance and improving model stability. After normalization, the categorical data labels were transformed into numerical format using Label Encoding, which assigns a unique integer to each fault class. Let $\vec{y} = \{y_1, y_2, \ldots, y_N\}$ represent the set of original categorical class labels. The label encoding transformation $\vec{y}'$ can be mathematically defined as:

$$\vec{y}' = B(\vec{y}), \text{ where } B : \mathbb{C}A \to \{0, 1, \ldots, J-1\}. \tag{24}$$

Where, $A$ is a set of categorical labels (e.g., PD and D); $J$ is the total number of classes; and $B$ is a deterministic mapping function that assigns a unique integer to each class.

To facilitate multi-class classification under the OVR training strategy, the encoded labels are further transformed into binary labels for each class $a \in \{0, 1, \ldots, J-1\}$, defined as:

$$y_r^{(a)} = \begin{cases} 1, & \text{if} y_r = a \\ 0, & \text{otherwise} \end{cases}. \tag{25}$$

This binarization enables the construction of $J$ binary classifiers, where each classifier is trained to distinguish one class from the rest. This encoding approach ensures consistent preprocessing across all models, facilitating efficient training and evaluation.

## 4. Implementation and results

This section presents the experimental setup, training process, and evaluation results of the proposed quantum classification framework. Performance metrics including accuracy, precision, recall, and F1-score are reported and visualized. The findings are discussed in the context of quantum machine learning capabilities, highlighting both the effectiveness and current limitations of the proposed approach.

### 4.1. Experimental configuration by VQC model

The proposed model for transformer fault diagnosis begins with data preprocessing, followed by the construction of the quantum model, including the feature map, ansatz, and optimization routine. Each class is then treated as an independent binary task, for which a dedicated VQC is trained. During inference, the outputs of all binary classifiers are aggregated to determine the final class label. The algorithm concludes with the computation of standard evaluation metrics to assess overall classification performance. (as depicted in Fig. 8, Fig. 9, and Algorithm 1).

**Algorithm 1.** OVR VQC Classification

Input: $\vec{x}=[T_x,T_y,P_x,P_y]$, $\vec{y}=\{y_1,y_2,...,y_N\}$
Output: Trained classifiers and classification metrics (accuracy, precision, recall, F1-score)
Step 1: Preprocessing
1: Input vector $\vec{x}=[T_x,T_y,P_x,P_y]$ normalize into $\vec{x}'=[T_x',T_y',P_x',P_y']$
2: Encode the categorical labels $\vec{y}$ using Label Encoding into $y'$
3: Split the dataset into training and test sets: $(\vec{x}'_{train},\vec{x}'_{test},\vec{y}'_{train},\vec{y}'_{test})\leftarrow$train_test_split$(\vec{x}'_i,\vec{y}')$
Step 2: VQC Configuration
4: Set number of qubits: $n=2$
5: Define feature map:
6: def parameterized_feature_map (num_qubits = 2, feature_dimension=4):
7: params = ParameterVector ('theta', feature_dimension)
8: qc = QuantumCircuit(num_qubits)
# Use gates $R_{YY}(\theta)$ and $R_{ZX}(\theta)$ to encode data.
9: qc.ryy(params[0], 0, 1) # $R_{YY}(T_x')_{0,1}$
10: qc.rzx(params[2], 1, 0) # $R_{ZX}(P_x')_{1,0}$
11: qc.ryy(params[1], 1, 0) # $R_{YY}(T_y')_{1,0}$
12: qc.rzx(params[3], 0, 1) # $R_{ZX}(P_y')_{0,1}$
13: return qc
14: $n=2$
15: feature_map=parameterized_feature_map(num_qubits = $n$, feature_dimension=4)
16: Define ansatz:
17: ansatz = EfficientSU2(num_qubits = $n$, entanglement='full', reps=4)
18: Select optimizer:
19: optimizer = COBYLA(maxiter=100)
Step 3: Training Phase (OVR)
20: Initialize empty dictionary classifiers $\leftarrow\{\ \}$
21: For each label L in unique_labels:
22: Create binary label: $y_{binary}\leftarrow(y'_{train}==\text{L})$
23: Initialize VQC: vqc ← VQC(optimizer, feature_map, ansatz)
24: Train VQC: vqc.fit( $x'_{train}$ , $y_{binary}$ )
25: Store classifier: classifiers $[\text{L}]\leftarrow$vqc
Step 4: Prediction Phase
26: def predict_OVR( $x'_{test}$ ):
27: Initialize empty list: predictions ← []
28: For each sample $x$ in $x'_{test}$ :
29: Initialize votes $\leftarrow\{\ \}$
30: For each label L in classifiers:
31: votes $[\text{L}]\leftarrow$classifiers$[\text{L}]$.predict( $x$ )
32: If any label L has votes$[\text{L}]==1$:
33: Assign predicted label ← first label with vote = 1
34: Else:
35: Assign predicted label ← label with max vote score
36: Append predicted label to predictions
37: return predictions
Step 5: Performance Evaluation
38: Compute predictions: $\hat{y}$ = predict_OVR( $x'_{test}$ )
39: Evaluate classification metrics: Accuracy, Precision, Recall, F1-score.

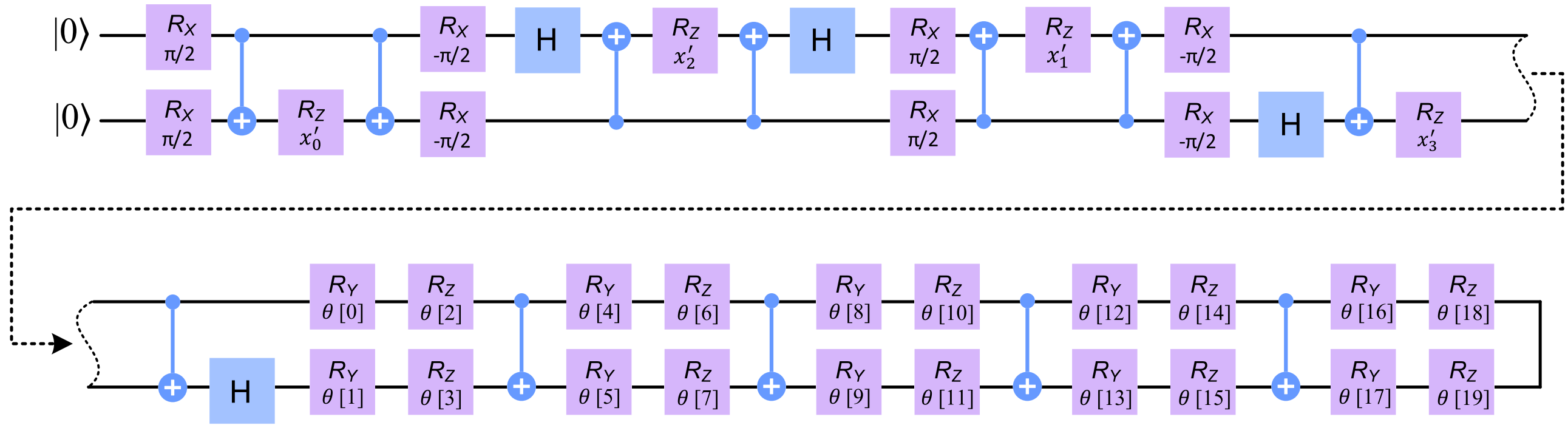

**Fig. 9.** A proposed VQC with customized feature map and four layers EfficientSU2 ansatz.

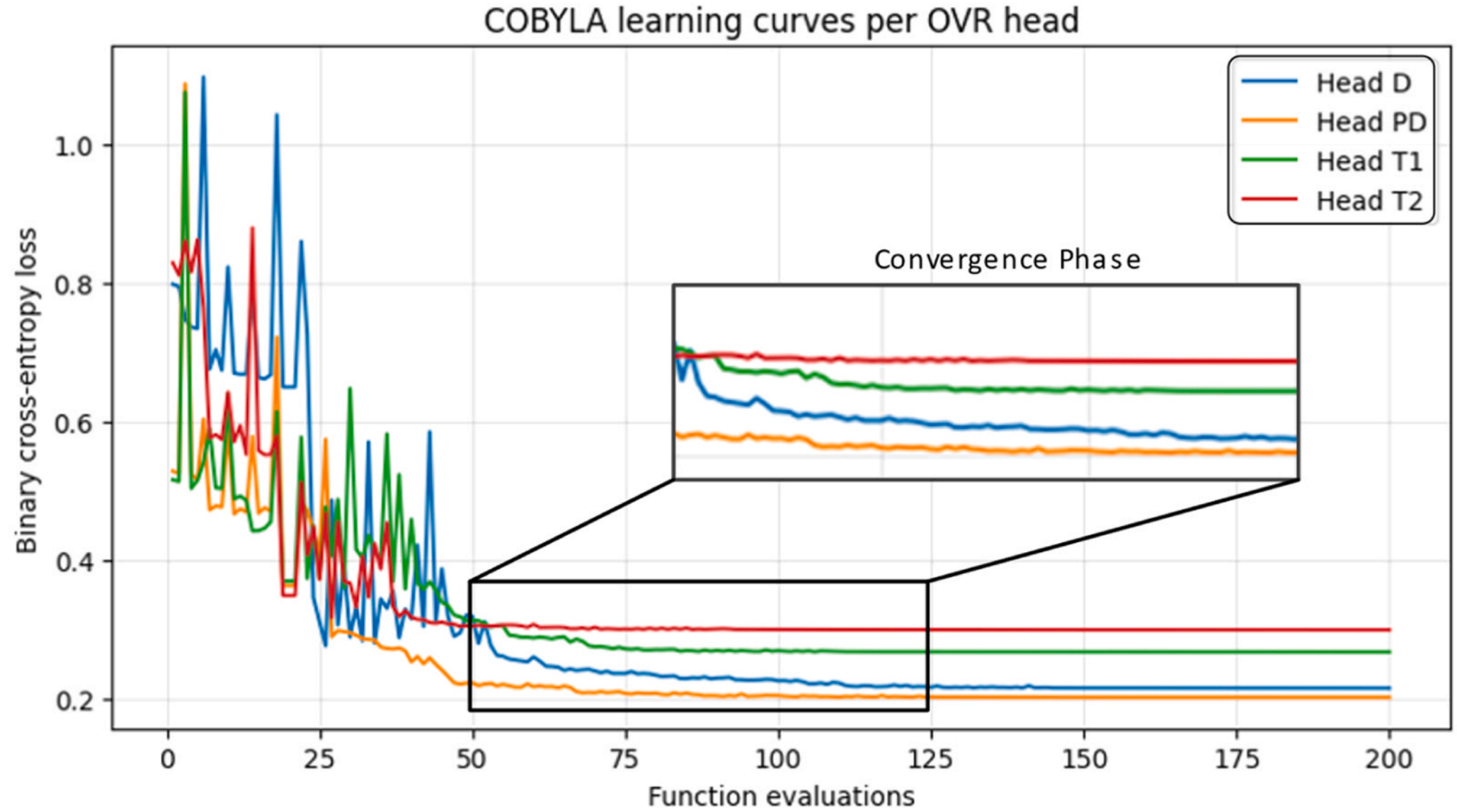

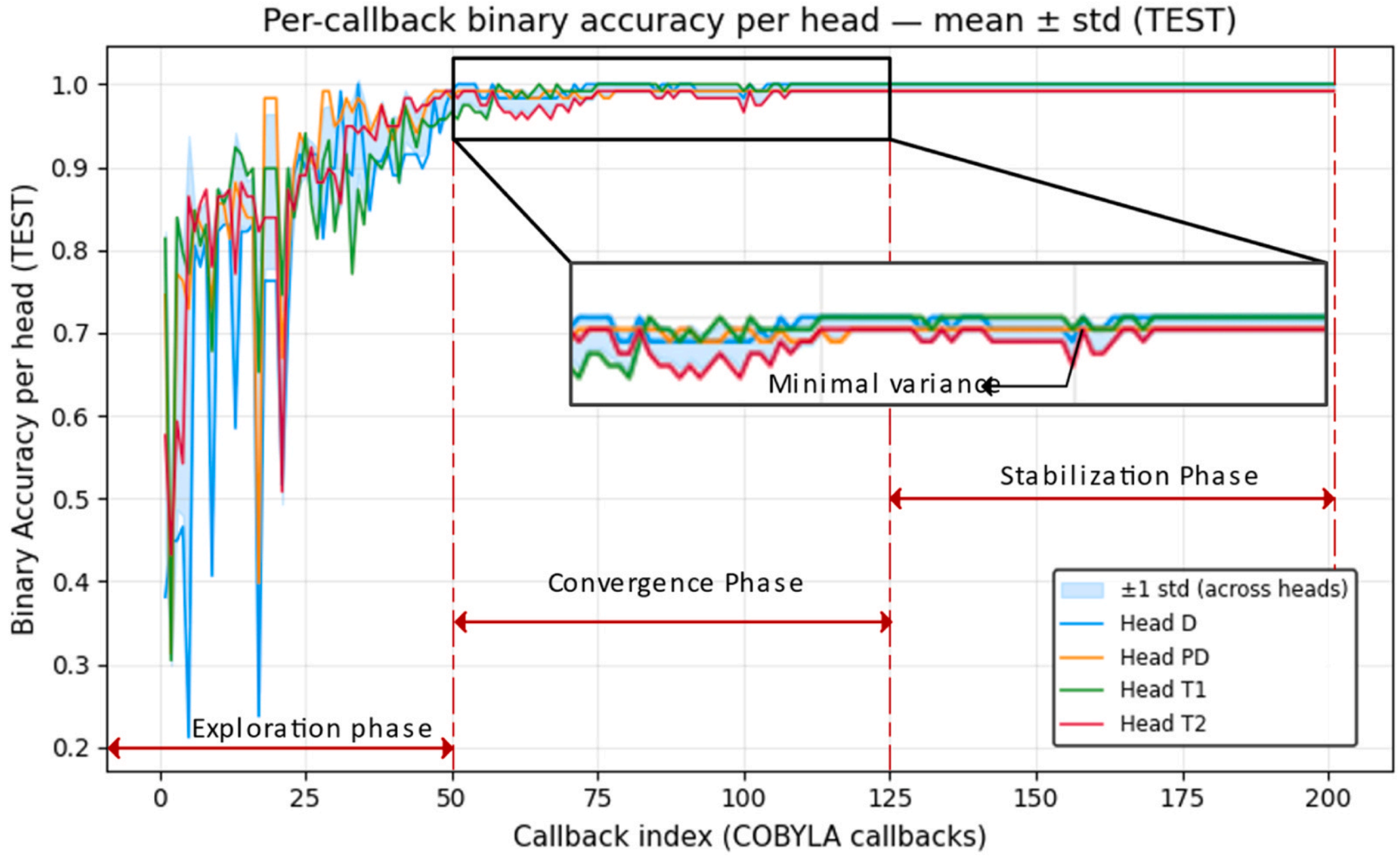

**Fig. 10.** Overall convergence behavior of the proposed COBYLA-based variational model.

### *4.2. Evaluation of the VQC model*

Subsection 4.2 presents a concise, tripartite evaluation of the VQC model: *(i)* an analysis of the feature space with cross-dataset validation, *(ii)* ablation studies on the feature map, ansatz, and optimizer to assess accuracy, convergence, and stability, and *(iii)* robustness assessments under simulated and real quantum noise. Collectively, these experiments identify a configuration that optimally balances accuracy, generalization, and hardware realism.

#### 4.2.1. Experimental result

Fig. 10 illustrates the convergence behavior of the proposed VQC. The global loss decreases smoothly and stabilizes toward convergence, while all OVR heads follow consistent trajectories, demonstrating that COBYLA effectively optimizes the variational parameters across classifiers.

This convergence is directly reflected in the evolution of the quantum state embeddings shown in Fig. 11. Before training, when only the feature map is applied (Fig. 11a), the samples are broadly dispersed on the Bloch sphere, exhibiting limited class separability despite the expressiveness of the ZX–YY encoder. After optimizing the EfficientSU2 ansatz (Fig. 11b), the embeddings become significantly more structured: class clusters contract, overlap reduces, and clearer inter-class boundaries emerge. This transition from the diffuse pre-training distribution to the organized post-training geometry confirms that the variational layers successfully reshape the Hilbert-space representation into a form that enhances decision boundaries.

Together, Fig. 10 and Fig. 11 provide consistent evidence that the training process is stable and that the VQC effectively transforms the initial feature-map embedding into a highly discriminative quantum representation.

To understand where the single misclassified T2 sample originates, we combine three complementary diagnostic views of the classifier's behavior. Each visualization captures a different aspect of the decision process: confusion matrix (Fig. 12), decision confidence (Fig. 13), and quantum-state geometry (Fig. 14). Together, they provide a coherent picture of how the model interprets this borderline case.

The confusion matrix offers the highest-level perspective by summarizing class predictions. It shows that only one T2 instance is assigned to PD, indicating that the error is isolated and class-specific. To examine why this occurs, we evaluate the decision margins produced by the OVR heads. Margin plots quantify how strongly the model prefers one class over another, and they reveal that the misclassified T2 sample lies extremely close to the PD boundary. Its margin falls just on the PD side, indicating that the classifier views this sample as nearly indistinguishable from PD in terms of decision confidence.

To probe this behavior from the quantum standpoint, we inspect the 3D correlator space (Fig. 14) derived from expectation values of the Pauli observables $\langle Z_0 \rangle$, $\langle Z_1 \rangle$, and the two-qubit correlator $\langle Z \otimes Z \rangle$. These measurements define a geometric embedding of the encoded quantum states. When plotted, the misclassified T2 sample appears not within the typical T2 region but instead at the edge of the PD cluster—its correlator tuple aligns more closely with PD than with other T2 points. This shows that, after encoding and variational transformation, the quantum state of this sample naturally occupies a region shared with PD patterns.

Taken together, these three diagnostic views, classification outcome, margin-based confidence, and quantum correlator geometry, independently identify the same instance as the outlier. The misclassification is therefore not a training defect but a consequence of the sample's intrinsic proximity to PD both in classical DGA space and in the quantum measurement space.

To move beyond a single random split, we adopt a cross-site evaluation protocol, leveraging two datasets that differ in provenance, oil-handling practices (Fig. 15). This setup introduces a realistic domain shift while remaining comparable to IEC-based prior work. All preprocessing steps, such as feature scaling and hyperparameter tuning, are applied strictly within the training set to avoid leakage. Crucially, rather than relying on loss-based class weighting or synthetic oversampling techniques (such as SMOTE or ADASYN) to handle the significant class imbalance (e.g., a ~4.5:1 ratio between class D and class PD), the proposed framework addresses it structurally through the One-vs-Rest (OVR) decomposition. By training a dedicated, independent binary quantum circuit for each fault class, the OVR strategy ensures that all minority classes receive specialized quantum feature encoding and variational ansatz optimization, effectively eliminating the risk of majority-class dominance.

This structural advantage is further amplified by the ZX–YY quantum feature map, which utilizes non-commuting rotations to generate pairwise interaction terms, creating highly separable cluster structures in the Hilbert space even for minority classes. Additionally, the extremely compact parameter budget (only 20 trainable parameters per head) combined with the derivative-free COBYLA optimization provides an implicit regularization effect, preventing the model from overfitting to the skewed class distribution. By optimizing dedicated quantum circuits per class, the proposed VQC provides a more principled approach to handling class imbalance for small DGA datasets, achieving higher accuracy than SMOTE-augmented models without the need to artificially expand the training distribution.

Under this protocol, to rigorously assess the model's reliability and ensure statistical significance on the independent IEC TC 10 test set, we calculate the Bootstrap 95% confidence intervals (CI) using B= 100,000 resamples. The VQC maintains highly stable generalization across all metrics, achieving Accuracy = 99.15% (95% CI: 97.46%–100.00%), Precision = 99.23% (95% CI: 97.46%–100.00%), Recall = 99.15% (95% CI: 97.15%–100.00%), and F1-score = 99.12% (95% CI: 97.03%–100.00%) (Fig. 15). Compared with the baseline configuration without the processing layer, the proposed model shows consistent gains across all metrics, demonstrating competitiveness with state-of-the-art methods and robust generalization to a distinct data source (see Subsection 4.4).

#### 4.2.2. Impact of feature map, ansatz, and optimizer design on VQC performance

To compare the expressivity of different feature maps, we compute their frame potential values (as detailed in Appendix A). The results show that our proposed compact feature map achieves the highest expressivity among all candidates. Fig. 16 further visualizes the geometric embeddings generated by each feature map. While conventional maps often restrict data to narrow submanifolds—frequently forming great-circle–like trajectories or producing sparse, weakly separable clusters—our feature map yields a significantly richer and more diverse embedding. The resulting distribution covers a broader region of the Bloch sphere, demonstrates greater curvature variability, and reveals more distinct inter-class separation. This enhanced spreading in the higher-dimensional quantum Hilbert space indicates stronger representational power, ultimately improving class separability for the downstream VQC.

Table 3 provides a comparative analysis of the classification accuracy of the VQC across different ansatz structures and varying repetition layers. Among the ansatz types evaluated, EfficientSU2 consistently surpasses both RealAmplitudes and PauliTwoDesign at all repetition levels. Notably, EfficientSU2 achieves an accuracy of 85.59% with a single layer and reaches a maximum of 99.15% with four repetitions.

Table 4 show, across optimizers, the main driver of wall-time is the number of circuit evaluations per iteration. Gradient-based methods with parameter-shift (Adam, RMSProp, SGD) require $\approx 2d$ circuit calls per step, so they are $> 50\times$ slower in our setting ($\approx$ 400 min median) yet converge to substantially worse test accuracy ($\approx$ 68–71%).

#### 4.2.3. Simulation-based robustness analysis under NISQ constraints

To further evaluate the robustness of the proposed VQC model under realistic quantum noise, we simulate depolarizing errors using Qiskit's noise modeling tools. Specifically, a NoiseModel is constructed with single-qubit depolarizing noise applied to all quantum gates. Depolarizing noise is a standard abstraction for quantum decoherence, where, with a given probability, the state of a qubit is replaced by a completely mixed state.

In this study, we apply depolarizing noise to all single-qubit gates (u1, u2, u3) with error probabilities of 1%, 5%, and 10%, respectively. The noise model is integrated into the AerSimulator backend to emulate

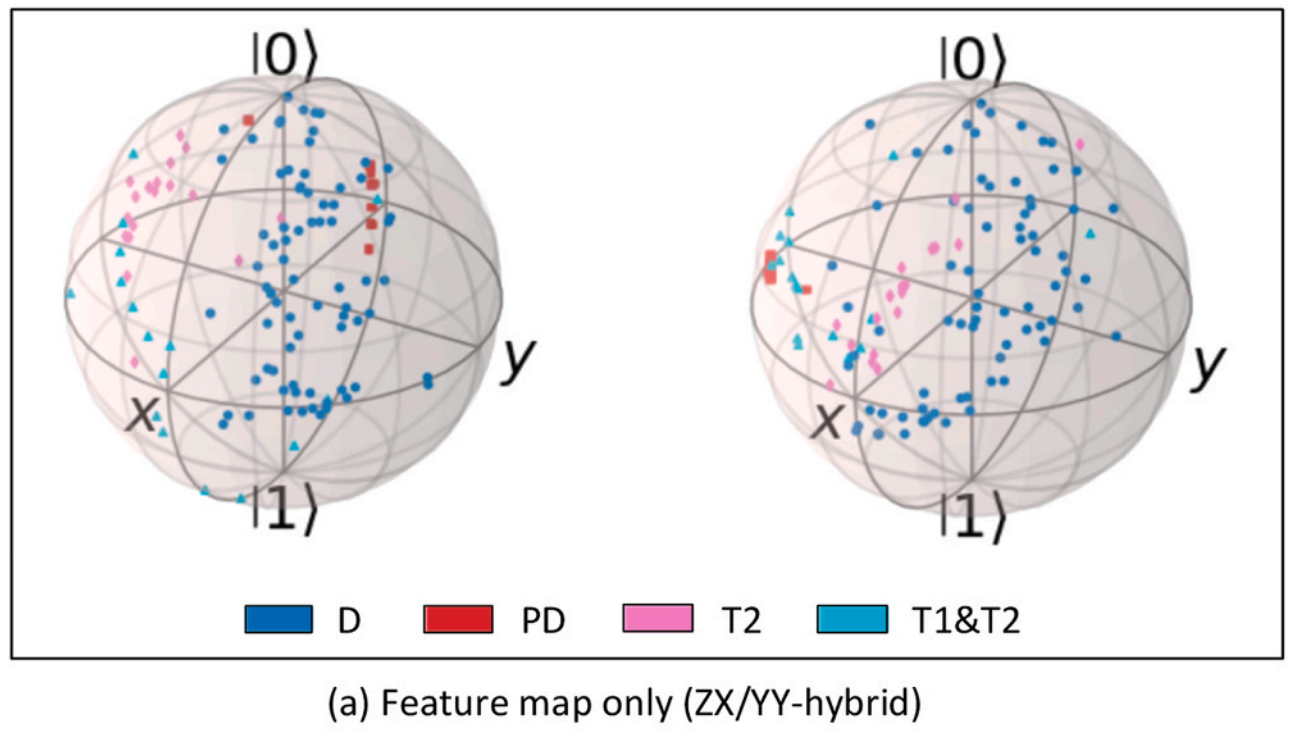

(a) Feature map only (ZX/YY-hybrid)

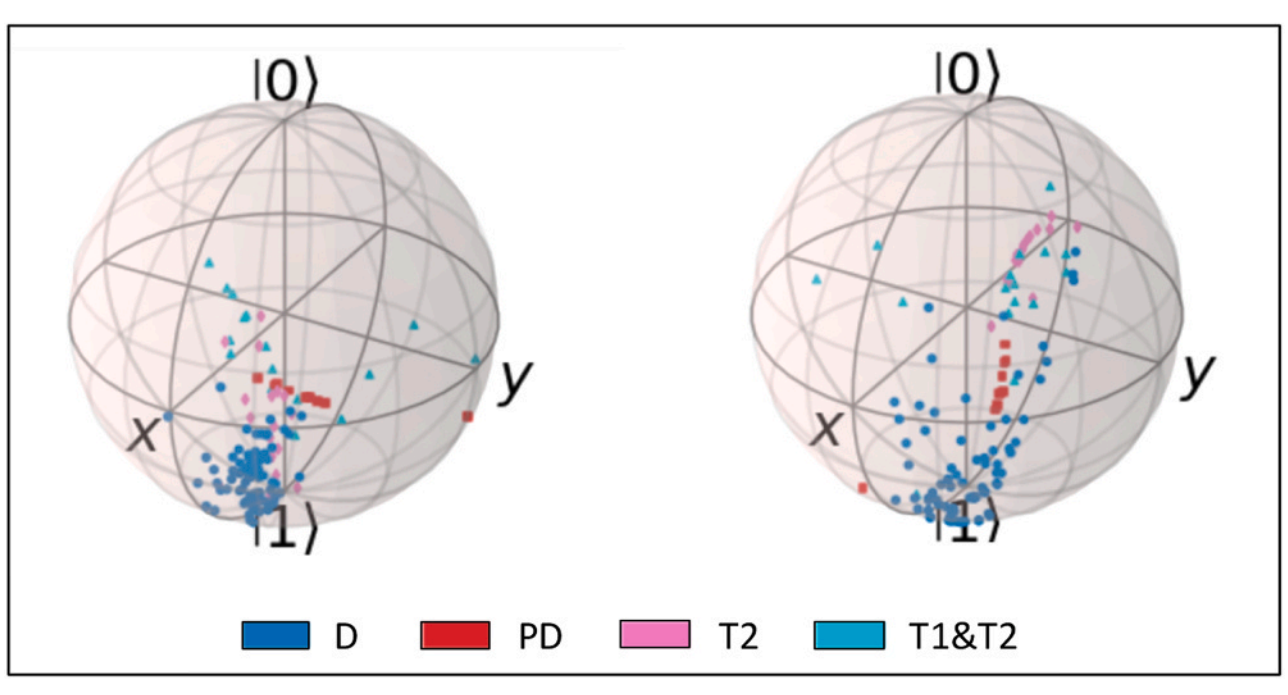

(b) Full pipeline (feature map + EfficientSU2 ansatz)

**Fig. 11.** Bloch-sphere visualization before and after ansatz training on the test dataset.

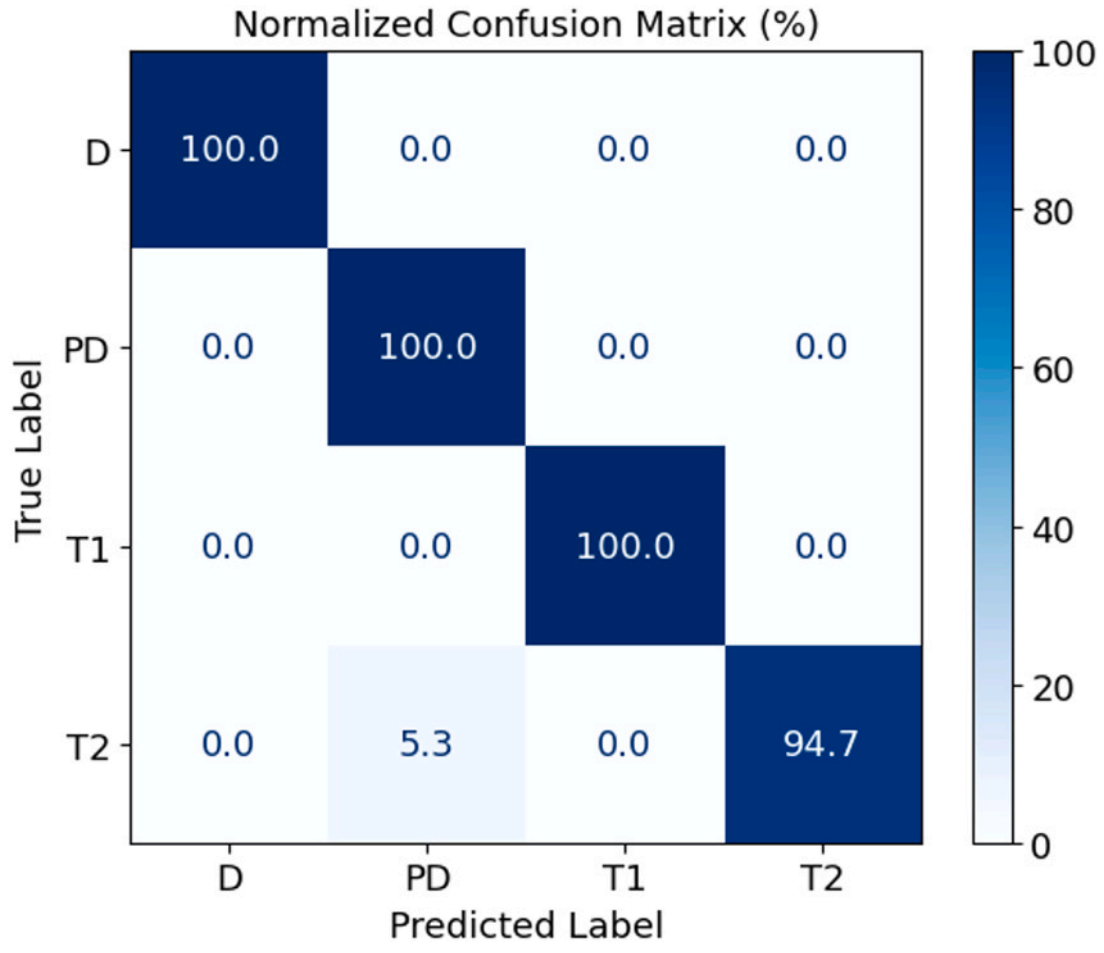

**Fig. 12.** Normalized confusion matrix of the quantum classifier on the test set.

quantum circuits under various error rates. The corresponding Qiskit algorithm (Algorithm 2) is shown below:

**Algorithm 2.** Noise model generator

```
from qiskit_aer.noise import depolarizing_error, NoiseModel
from qiskit_aer import AerSimulator
def get_backend_with_noise(error_rate):
    noise_model = NoiseModel()
    error = depolarizing_error(error_rate, 1)  # single-qubit error
    noise_model.add_all_qubit_quantum_error(error, ['u1', 'u2', 'u3'])  # apply to all gates
    return AerSimulator(noise_model=noise_model)
```

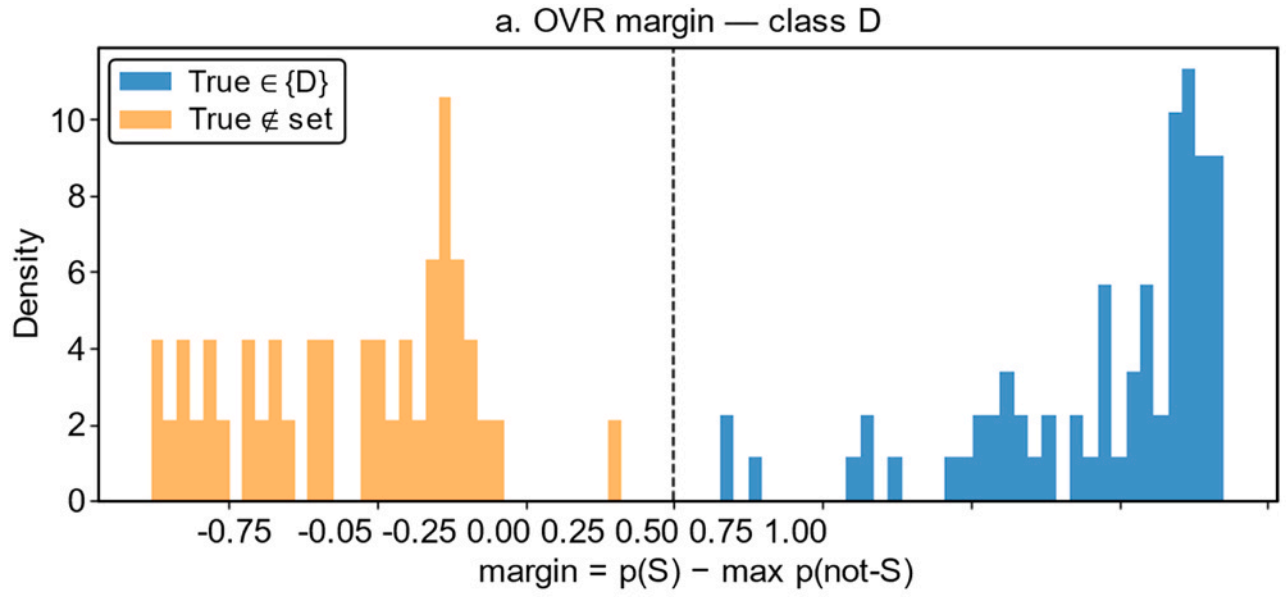

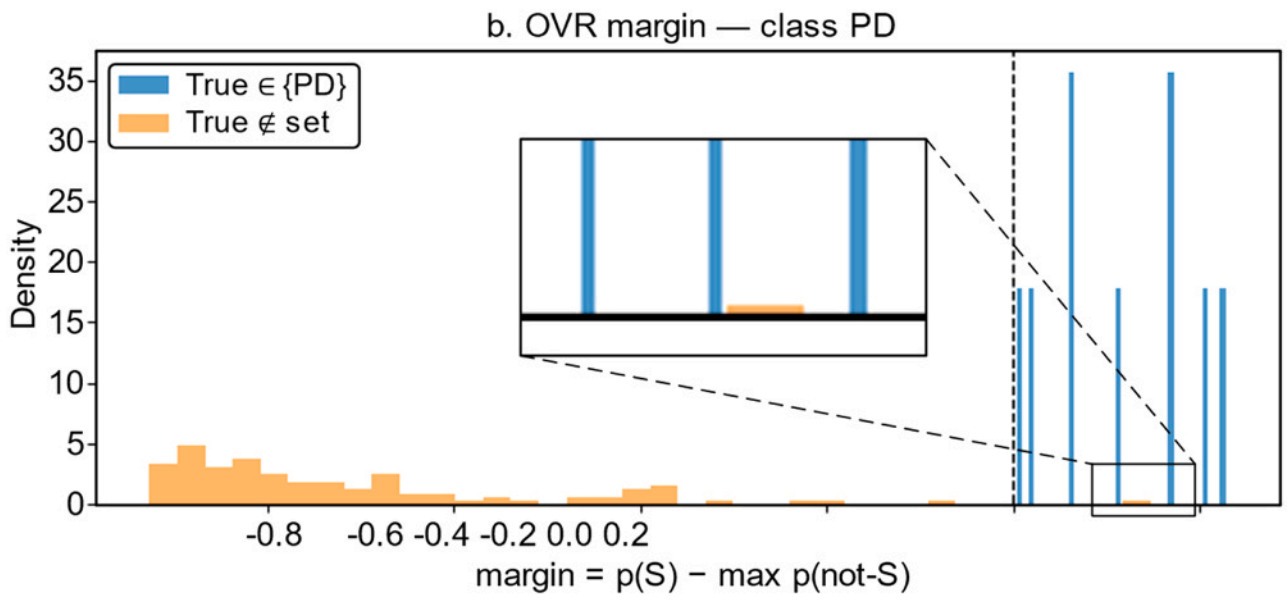

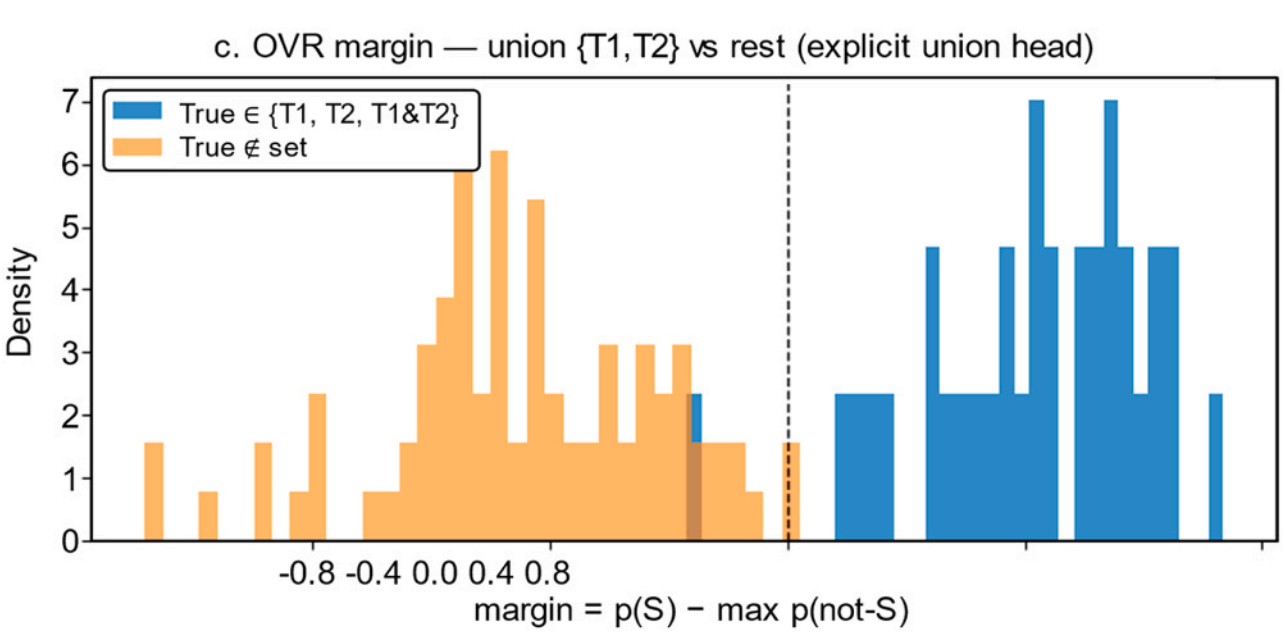

**Fig. 13.** One-vs-Rest (OVR) margin distributions on the test dataset.

By adjusting the error_rate parameter, we emulate increasing levels of depolarizing noise to approximate NISQ hardware conditions and assess the model's robustness. As summarized in Table 5, the VQC remains stable under moderate noise: performance decreases only slightly from 1% to 10%, reflecting the expected impact of decoherence on state fidelity. Low noise levels produce negligible degradation, indicating that both the circuit structure and optimization process are resilient to minor hardware errors. Even at 10% noise, the model still achieves 95.76% accuracy and an F1-score of 0.95, demonstrating that it can maintain meaningful decision boundaries under noisy quantum operations. Training time remains comparable across noise settings, suggesting that convergence is not significantly affected and supporting the feasibility of deploying the model on NISQ devices.

We also conducted a limited shots experiment utilizing the trained VQC heads (feature map + EfficientSU2, reps = 4) on an IBM backend (ibm_brisbane, 8192 shots, light readout-mitigation enabled) (Fig. 17). The 50 test samples were drawn sequentially from the IEC TC 10 dataset

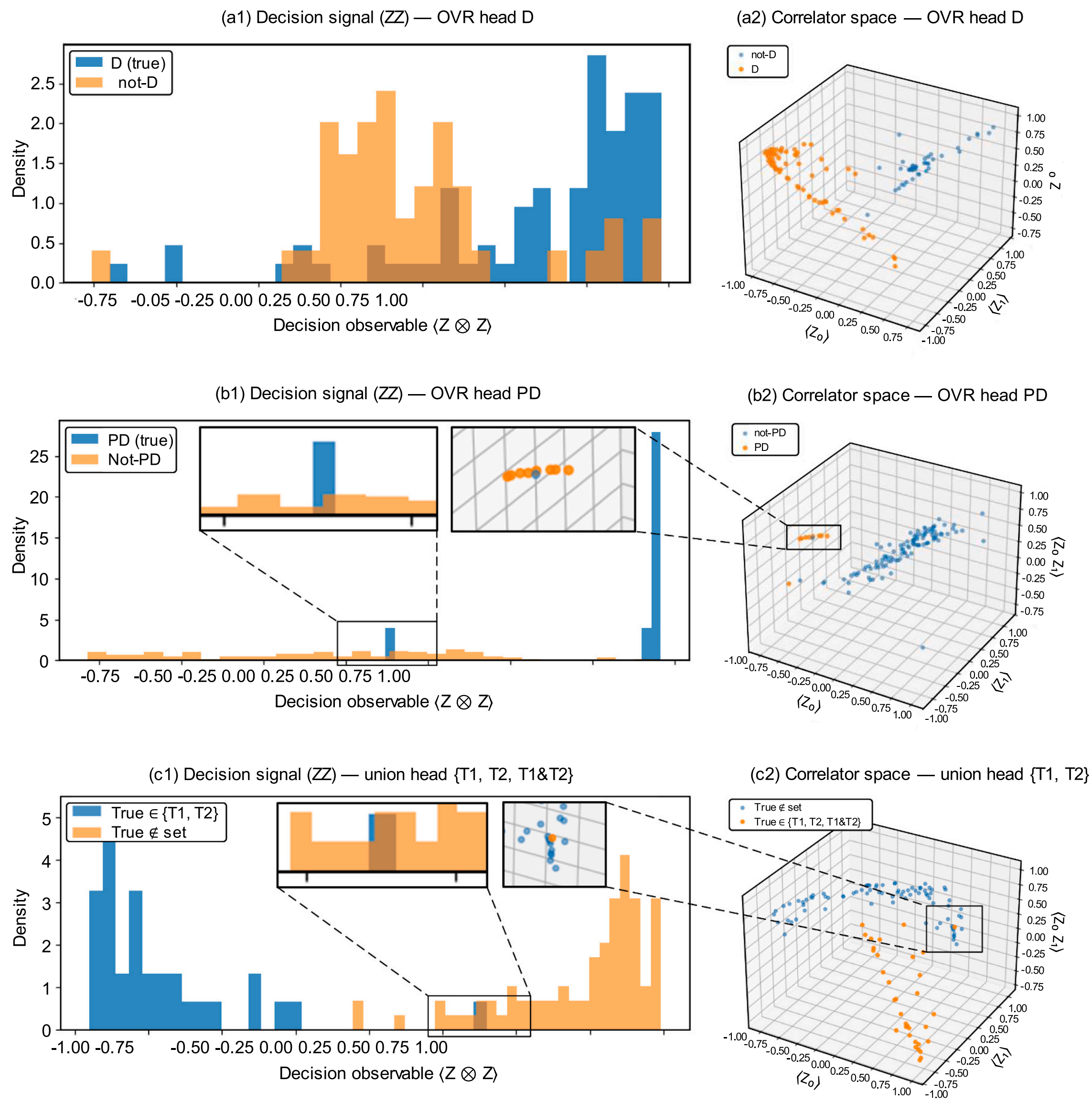

**Fig. 14.** Decision observables and correlator space visualization for OVR heads on the test dataset.

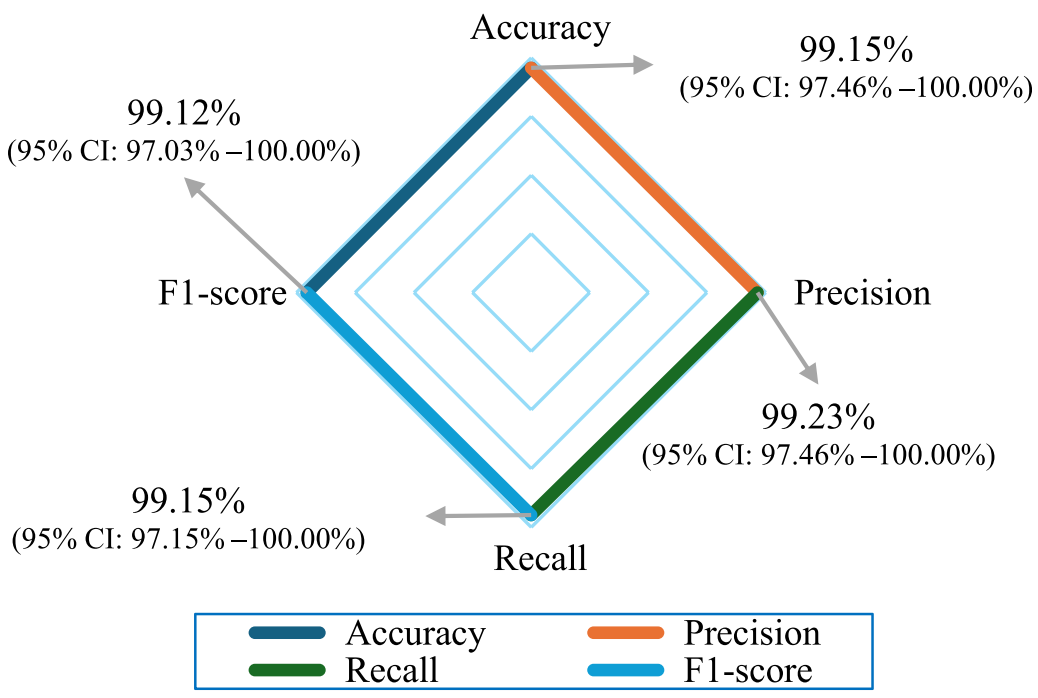

**Fig. 15.** Evaluation metrics of the model on the test set (accuracy, precision, recall, F1-score).

[48] in its original ordering, primarily constrained by the cost of quantum hardware execution on the IBM public plan (billing ≈ $1.6/s). Regarding hardware inference efficiency, the average time on the real backend is approximately 4.8 s/sample, which includes queue waiting time and shot accumulation. However, the pure per-circuit execution time, excluding the queue, is remarkably fast at roughly 0.15 s. On this held-out subset of 50 test samples, the hardware execution achieved an accuracy of 97.99% (95% CI 94.00%–100.00%), B = 100,000 resamples. Compared to the exact statevector simulator which yielded 100.0% on the same subset, this represents a minimal performance drop of $\Delta \approx 2.0$ %age points.

To quantify agreement at the probability level we compare $p_1 = (1 - \langle ZZ \rangle)/2$ per circuit:

- Mean absolute deviation $E\left[\left|p_1^{HW} - p_1^{Sim}\right|\right] = 0.0294$; median $= 0.0298$; max $= 0.0707$.
- The scatter plot $p_1^{Sim}$ vs $p_1^{HW}$ clusters tightly around the diagonal, indicating near-linear agreement across the full dynamic range.

These numbers are consistent with shallow two-qubit depth and the backend's reported median readout error (~2–3%), and they confirm that our $ZX/YY$ encoder + SU2 ansatz is hardware (HW)-robust: the ranking of predictions is preserved and the end-to-end accuracy loss stays within a small constant (< 2–3 pp) without any circuit redesign. For further information regarding the deviations in probability distributions between hardware and simulation, please refer to Fig. 18. For details on the scatter of probability, please consult Fig. 19.

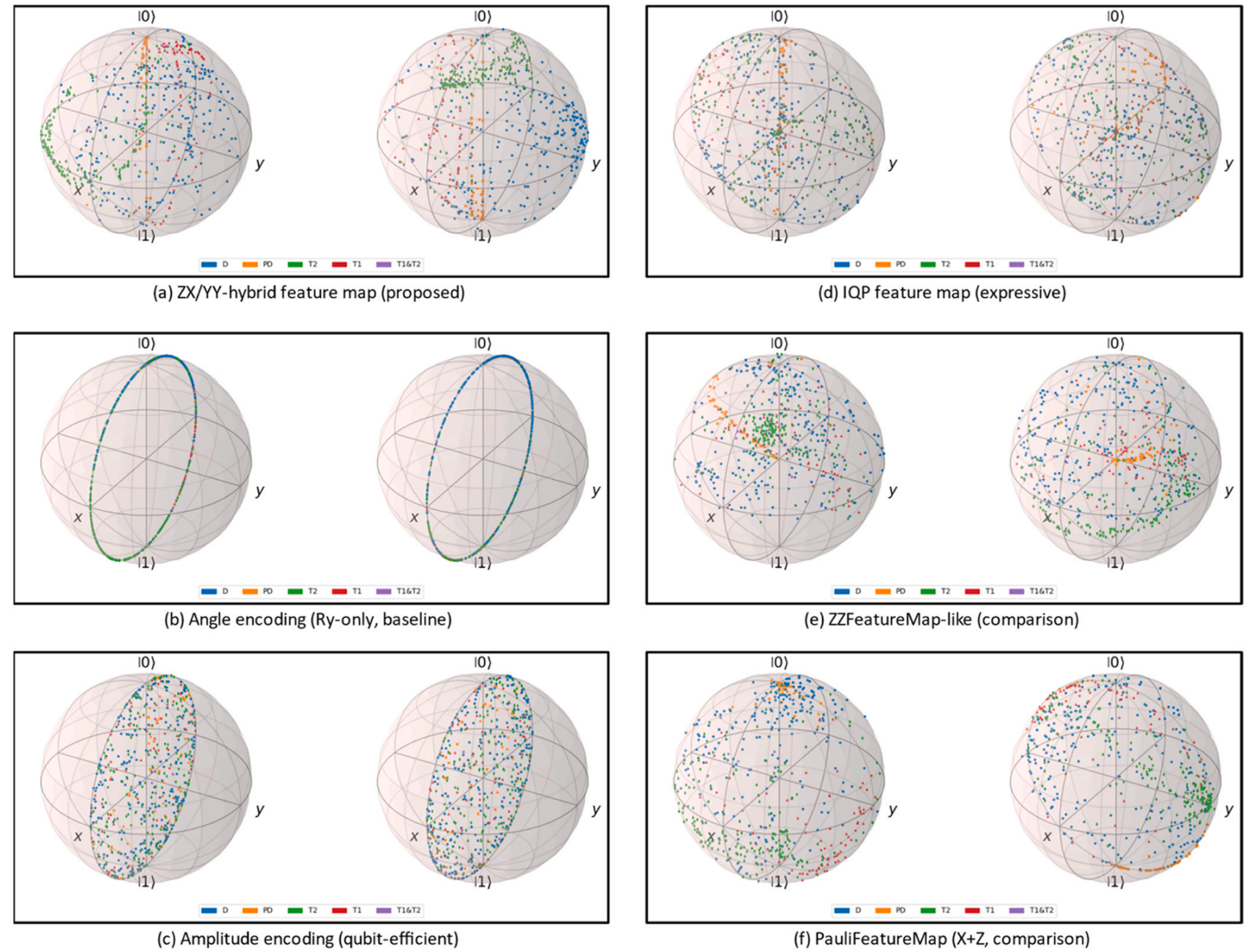

**Fig. 16.** Bloch-sphere visualization of the full dataset after quantum feature mapping.

**Table 3**
Comparative results of VQC accuracy with varying ansatz structures and repetition layers.

| Ansatz Type | One-layer (reps=1) | Two-layer (reps=2) | Three-layer (reps=3) | Four-layer (reps=4) |
|---|---|---|---|---|
| EfficientSU2 | 85.59% | 94.92% | 97.46% | 99.15% |
| RealAmplitudes | 72.88% | 83.90% | 83.90% | 83.90% |
| PauliTwoDesign | 50.00% | 72.88% | 77.12% | 83.90% |

### 4.3. Evaluation of computing resources

Table 6 compares the computational efficiency of the proposed VQC with existing QML approaches. A clear trend emerges: although models such as VQSL [29] and ATQCNN [52] offer strong expressivity, their training times exceed 400–600 min even on high-performance CPUs or GPUs, making them impractical for real-time or resource-limited settings. In contrast, the proposed VQC achieves a training time of only 7.36 min on a standard AMD 5400 U CPU using Qiskit's AerSimulator, demonstrating a substantial reduction in computational burden. This efficiency extends to inference, requiring only 2.3 ms per sample across all four OVR heads, demonstrating its practicality for real-time monitoring settings. While the classical ML baseline remains the fastest (0.11 min), it lacks the representational capacity of quantum models. Overall, these results show that the proposed VQC strikes a favorable balance between efficiency and expressivity, narrowing the practicality gap between classical and quantum methods and supporting deployment on near-term NISQ hardware.

**Table 4**
Comparison of existing optimizers.

| Optimizer | Type | Per-iter circuit evals | Train time (median) | Accuracy |
|---|---|---|---|---|
| COBYLA (chosen) | Derivative-free (simplex) | low (objective only) | ~ 7.36 min | 99.15 |
| Adam (parameter-shift) | First-order (adaptive gradient) | high (≈2devals/iter) | ~ 416.61 min | 71.36 |
| RMSProp (parameter-shift) | First-order (adaptive gradient) | high (≈2d evals/iter) | ~ 427.29 min | 68.35 |
| SGD / SDG (parameter-shift) | First-order (adaptive gradient) | high (≈2d evals/iter) | ~ 395.63 min | 67.98 |
| SPSA | Zeroth-order (stochastic) | medium (2 evals/step) | ~ 126.17 min | 75.26 |
| Nelder–Mead | Derivative-free (simplex) | low (objective only) | ~ 9.26 min | 71.27 |

**Table 5**
Impact of depolarizing noise levels on proposed VQC model performance.

| Noise's level | Accuracy (%) | F1-score | Training time (minute) | Description |
|---|---|---|---|---|
| Ideal (0%) | 99.15% | 0.99 | 7.36 | No noise added; represents theoretical upper-bound performance. |
| Depolarizing 1% | 98.30% | 0.98 | 7.16 | Slight performance drop; VQC remains robust under low noise. |
| Depolarizing 5% | 97.46% | 0.97 | 8.12 | Moderate noise begins to affect convergence and accuracy. |
| Depolarizing 10% | 95.76% | 0.95 | 7.75 | Accuracy degradation is more visible; training is still stable but less effective. |

The improved VQSL framework [29] incorporates classical fully connected neural network (FCNN) layers as a deliberate design choice to enhance interpretability and post-processing flexibility. When properly accounting for both the per-window local circuit repetitions and the corresponding FCNN parameters, the total trainable parameter count reaches 98 (Table 7), reflecting a hybrid design philosophy that prioritizes classical post-processing capacity. By contrast, the proposed VQC relies on a compact 20-parameter quantum-only classifier, demonstrating that comparable or higher accuracy can be achieved with a substantially smaller optimization landscape (20 trainable parameters versus 98 effective parameters in [29]).

Table 8 provides a side-by-side comparison between the improved VQSL framework [29] and the proposed VQC model. The results highlight the substantially lower resource requirements of the proposed approach—using only two qubits, fewer parameters, and a lightweight global ansatz—while achieving markedly higher accuracy (99.15% vs. 95.4%) and dramatically reduced training time. These differences underscore the computational efficiency and practical deployability of the proposed VQC relative to more complex QML architectures.

### *4.4. IEC TC 10 database validation and comparison with current techniques*

Table 9 presents the comparative diagnostic results across a wide range of rule-based, VQSL, and both classical and hybrid machine-learning and deep-learning baselines, together with the proposed VQC model. These results (see the columns "Reported accuracy (%)" and "Proposed VQC model accuracy (%)") clearly indicate that the VQC achieves superior diagnostic accuracy and stability compared with other methods. It should also be noted that the accuracy values of the proposed VQC model corresponding to the rows associated with references [60–63] were not provided due to the unavailability of data. To understand the origin of this performance advantage, the following analysis explains how the non-commuting ZX–YY encoder enhances feature entanglement and geometric separability within the Hilbert space.

Table 10 provides a qualitative window into the decision-making process of the proposed VQC by listing the four class-probabilities returned for ten representative TC 10 records.

The sole discrepancy across the full 118-sample test set arises in sample 10, where the VQC predicts PD (class-PD = 0.776) while the reference label is "T1 & T2." This disagreement is readily interpreted: $H_2$ reaches $2.31 \times 10^3$ ppm, well within the IEC 60599 partial-discharge band, whereas $CH_4$ and $C_2H_2$ remain low; consequently, a PD diagnosis is chemically plausible. The mis-match therefore reflects overlap in gas signatures rather than a systematic model failure. Excluding this borderline case, the VQC attains perfect identification on the remaining 117 records, and the probability outputs in Table 10 demonstrate that the quantum classifier not only achieves state-of-the-art accuracy but also provides transparent confidence measures that can be cross-checked against expert gas-ratio reasoning.

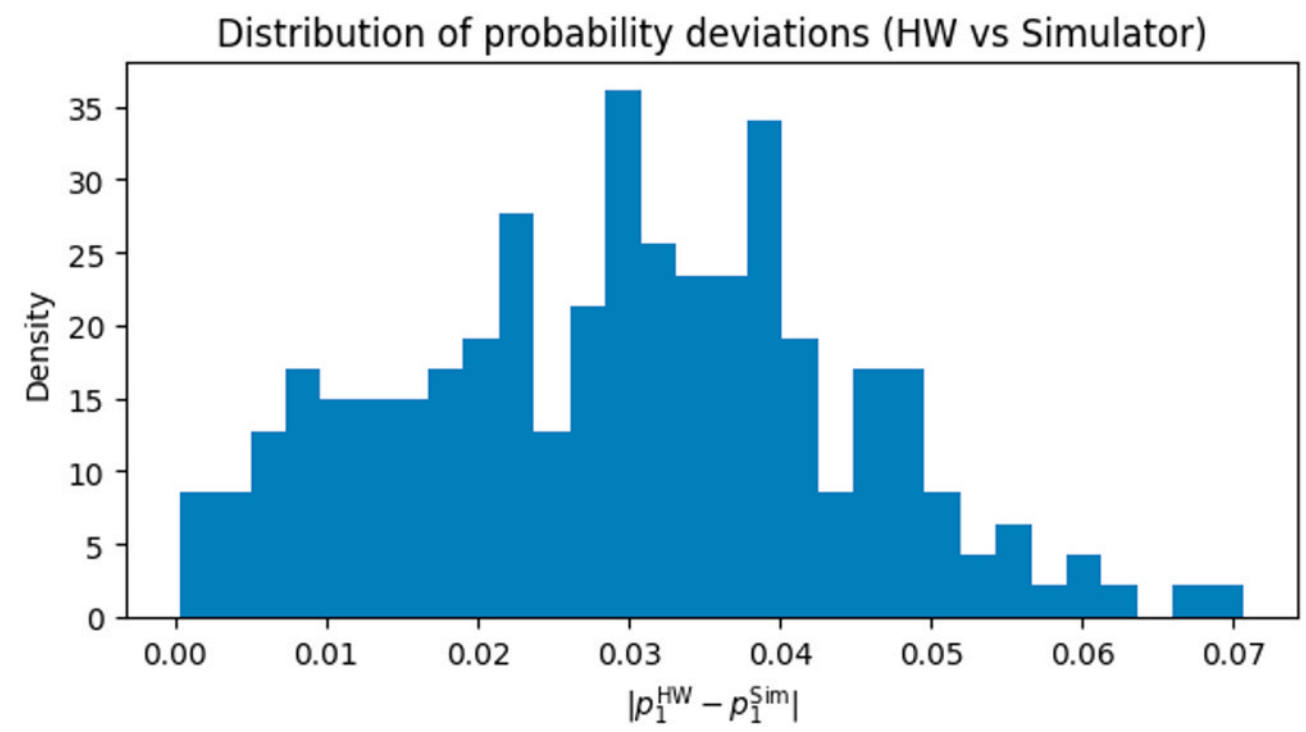

**Fig. 18.** Distribution of $|p_1^{HW} - p_1^{Sim}|$ across 120 pubs (50 samples × OVR heads) on *ibm_brisbane*. Mean = 0.0294, median = 0.0298, max = 0.0707.

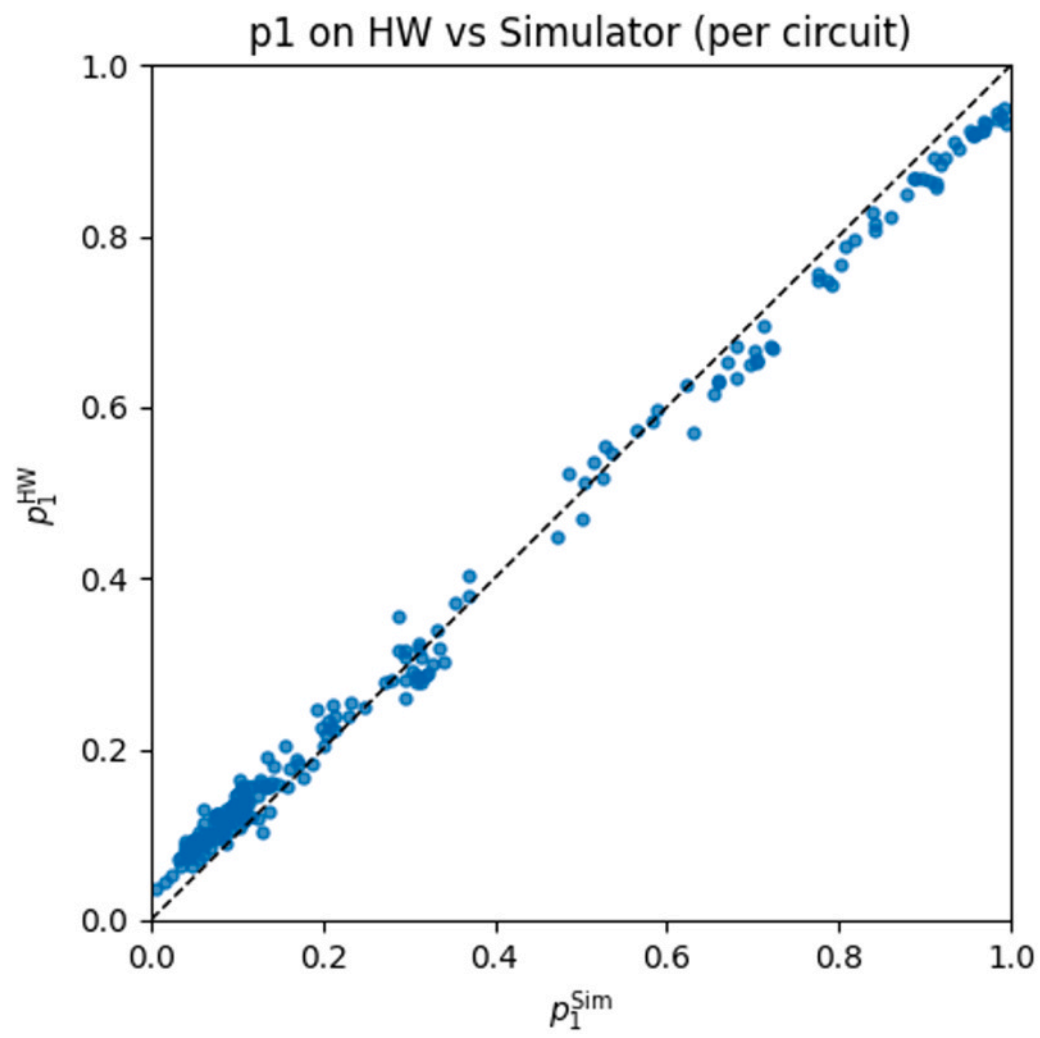

**Fig. 19.** Scatter of $p_1^{Sim}$(x − axis) vs $p_1^{HW}$(y − axis). Points lie close to the y = x line, showing strong agreement.

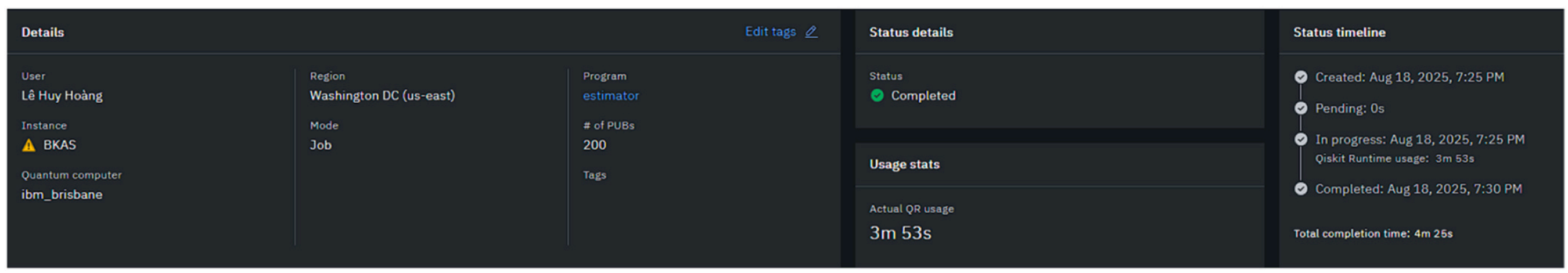

**Fig. 17.** Job status of the real ibm-brisbane backend.

**Table 6**
Computational efficiency comparison of the proposed model and existing QML methods.

| Ref | Method / Framework | Backend / hardware | Training time (minute) |
|---|---|---|---|
| [7] | Classical ML | Intel Core i5–7300HQ CPU (2.5 GHz) | 0.11 |
| [29] | VQSL | Intel Core i7-13700K NVIDIA GeForce RTX 4070 GPU | 443.885 |
| [51] | FH-QVC-DRC | AMD Ryzen7 CPU and 64 GB RAM | ~ 60.00 |
| [52] | ATQCNN | Intel Core i7–8700 CPU (3.2 GHz) and 32 GB RAM. | 673.54 |
| [53] | VSQL | Paddle Quantum Framework | 425.72 |
| **Proposed model** | **VQC (Qiskit)** | **AerSimulator (on AMD 5400 U CPU)** | **7.36** |

## 5. Discussion

The results in Sections 4.2–4.4 consistently demonstrate that the proposed ZX–YY hybrid feature map, combined with the compact VQC architecture, provides a structurally meaningful advantage for transformer fault classification. The use of non-commuting Pauli rotations introduces cross-feature interactions that create a richer and more separable embedding manifold than conventional angle-based or commuting quantum encoders. This enhanced representational geometry directly aligns with the strong diagonal dominance observed in the confusion matrix and the elevated accuracy, recall, and F1-scores reported in Fig. 15. These improvements are therefore attributable not to incidental optimizer behavior but to the expressivity induced by the feature-map design and its ability to capture nonlinear DGA relationships.

The experimental results further highlight the efficiency of the proposed VQC. With only 20 trainable parameters and a modest dataset of 588 samples, the model achieves a diagnostic accuracy of 99.15%, requiring significantly fewer data and computational resources than classical ML methods or prior hybrid QML approaches such as the improved VQSL. A closer analysis also clarifies that although [29] reports low parameter counts, its actual effective optimization load is substantially larger than reported due to additional classical neural layers not reflected in the reported parameter count, as well as repeated shadow sampling windows, resulting in more than 98 effective trainable parameters. In addition, a dimensional consideration arises in the application of amplitude encoding in [29], where 5 features are mapped onto 5 qubits ($2^5 = 32 \neq 5$), which does not satisfy the dimensionality requirements of amplitude encoding. These observations are presented not to diminish the contribution of [29], but to contextualize the methodological choices and parameter efficiency of the proposed VQC within a fair comparative framework.

Despite these strengths, several limitations remain. First, the dataset is relatively small and exhibits notable class imbalance, which constrains the statistical resolution of the evaluation and may amplify the influence of borderline samples, such as the single T2 instance that overlaps with PD in both classical and quantum embedding spaces. Second, most experiments rely on noiseless or lightly-noisy simulations; although small-scale hardware validation indicates reasonable robustness, the performance achievable on current NISQ devices remains sensitive to calibration drift, readout noise, and backend-specific compilation constraints. Third, while the OVR decomposition enables a practical multiclass extension of the binary VQC, it increases the number of required circuits and may propagate per-head calibration differences into the final decision.

To address the dimensional limitations of the current two-qubit design, a concrete scalability pathway can be established. The ZX–YY encoding naturally extends to higher qubit counts by interleaving additional two-qubit rotation blocks across qubit pairs. For $n$ qubits, the encoder can accommodate $2^n$ features with $n$ two-qubit gates per upload layer, maintaining the shallow-depth advantage. Future work will investigate 4–8 qubit circuits with multi-sensor DGA datasets that incorporate additional gas concentrations (CO, $CO_2$) and thermal parameters. Data re-uploading strategies, which repeat the encoding layer across multiple circuit depths, offer another route to increased expressivity without increasing qubit count.

**Table 7**
Detailed comparison of trainable parameters and quantum resources.

| Model | Global qubits | Trainable quantum parameters | Trainable classical parameters | Total parameters | Circuit depth | Feature-map / Ansatz |
|---|---|---|---|---|---|---|
| Proposed **VQC** (this work) | 2 | 2 qubits × 2 + 2 qubits × 2 × 4 layers = 20 | 0 (no post-classifier) | 20 | 4 | Custom 2-qubit hybrid feature map + EfficientSU2 (full entanglement) |
| Improved **VQSL** (Method 6, IEEE TPWRD 2025 [29]) | 5 (3-qubit sliding window) | 3qubits× 2$\theta$ /layer × 3layers = 18 | 7outputs×$C_5^3$inputs = 21weights+$C_5^3$bias = 80 | 18 + 80 = 98 | 3 | 5-qubit shadow feature extractor + FCNN output layer |

**Table 8**
Comparative analysis between the proposed VQC model and the improved VQSL method.

| Criteria | Improved VQSL (IEEE TPWRD 2025) (He et al., 2025) [29] | Proposed Model (VQC) |
|---|---|---|
| Number of Qubits | 5 global qubits + 3-qubit localized circuits | 2 qubits |
| Total parameters | Quantum_Params+FCNN_Params= 18 + 80 = 98 | 20 |
| Data Encoding | Amplitude encoding | Custom feature map (entanglement + angle-based) |
| Quantum Circuit Type | Localized variational circuits with parameter sharing | Fully entangled global ansatz (EfficientSU2) |
| Classification Strategy | Fully connected neural network (FCNN) with extracted shadow features | One-vs-Rest (OVR) using binary VQC models |
| Optimizer | Adam (gradient-based backpropagation) | COBYLA (gradient-free) |
| Training Backend | Paddle Quantum + NVIDIA RTX 4070 GPU | Qiskit + AerSimulator (CPU simulation on AMD 5400 U) – very weak hardware |
| Trainable Parameters | Reduced via localized design and shared parameters | Low (depending on reps and qubits) |
| Noise Resilience | Evaluated under noisy conditions (0%, 1%, 10%, 20%) | Simulated depolarizing noise (0%, 1%, 5%, 10%) |
| Achieved Accuracy | 95.4% (972 samples for training and 108 IEC TC 10 [48] samples for testing) | 99.15% (470 samples [47] for training and 118 IEC TC 10 [48] samples for testing) |
| Training Time | 443.885 min for 500 iterations, batch size 30 (GPU-accelerated) | 7.361 min for 4 OVR's Circuits (on CPU simulator) |

**Table 9**
Performance comparison of DGA models reported in the literature.

| Ref. | Model | Number of data samples | Reported accuracy (%) | Proposed VQC model accuracy (%) | Key limitations |
|---|---|---|---|---|---|
| [29] | VQSL | 108 | 95.4% | 99.1% | Parallel parameter-sharing local PQCs; robust in noise (≥93.8% at 20% depolarizing) |
| [32] | GA-SVM | 118 | 87.18% | 99.15% | Relies on hand-selected gas ratios; sensitive to hyper-parameter search. |
| [33] | ANFIS | 100 | 99.00% | 100% | High accuracy on a smaller subset; five parallel neuro-fuzzy models. |
| [34] | Hybrid CNN + SMOTE | 108 | 92.3% | 99.07% | Accuracy still depends on manual feature design; SMOTE balances raw data without fault-specific insight. |
| [35] | Hybrid DNN + B-SMOTE | 38 | 84.21% | 100% | B-SMOTE may reduce accuracy by generating synthetic samples near class boundaries; less effective for extremely rare classes. |
| [36] | Bridging Duval's method and DNN | 24 | 95.8% | 100% | DNN inherits Duval-based labels, so subjectivity and mislabeling can degrade its performance. |
| [37] | DADDNN | 118 | 92.9% | 99.15% | DADDNN still heavily depends on initial hyperparameter selection-especially learning rate—even with dynamic adjustment. |
| [38] | ADASYN | 103 | 87.17% | 100% | SMOTE/ADASYN can still bias minority-class decision regions. |
| [39] | Hybrid TLR - ADASYN + SO-RF | 117 | 93.98% | 99.14% | TLR-ADASYN requires multiple resampling rounds, increasing training time compared to single-pass methods. |
| [54] | Domain knowledge + CapsNet | 117 | 94.02% | 99.14% | Requires 25 handcrafted features and deep network (>200 epochs). |
| [55] | Modified new approach DGA | 99 | 71.71% | 100% | Strong feature overlaps between fault types leads to high confusion. |
| [56] | Hexagon method | 118 | 93.60% | 99.15% | Still rule-based; accuracy saturates when gas clusters overlap. |
| [57] | Improved three-ratio method | 113 | 81.42% | 99.11% | Original version only 71.5% accurate and suffers heavy fault-zone overlap; even the “modified” version tops out at 84.7%; purely empirical thresholds still miss mixed or evolving faults. |
| [58] | GALO-FIS RFRT & GALO-FIS IECRCT | 55 | 92.57% & 91.58% | 100% | Outperforms Roger and IEC methods but yields slightly lower accuracy, lacks stability analysis, and faces FIS scalability and integration limits. |
| [59] | AdaBoost-TCNN AdaBoost-WBLS SARIMA-CNN-GRU | 46 | 95.35% 93.02% 90.70% | 100% | Relatively low diagnostic efficiency (long time consumption) and the need for future research on imbalanced data processing |
| [60]* | IRF + MLP | 164 | 87.80% | – | The potential risk of data leakage when determining the IRF threshold using test data may have limited the model’s accuracy to 87.80%. |
| [61]* | FT-Transformer | 214 | 97.10% | – | Relies on Gaussian Copula’s normality assumption; unstable for abnormal data and less effective for rare faults. |
| [62]* | DNN + SMOTE-ENN | 324 | 98.19% | – | Requires long training (300 epochs with Adam optimizer) and high resources, using batch size 128 for DPM1 and 32 for DPM2. |
| [63]* | NRBO-XGBoost | 358 | 92.10% | – | High computational cost in feature selection and limited scalability due to heavy hardware requirements. |

Note: Asterisked references ( [60–63]) use datasets not publicly available for re-evaluation; their results are provided for reference only and are not directly comparable with the IEC TC 10,118-sample test set used in this study.

**Table 10**
Examples of typical diagnostic outputs from the proposed model.

| Sample | $H_2$ | $CH_4$ | $C_2H_6$ | $C_2H_4$ | $C_2H_2$ | Class (D) | Class (PD) | Class (T1) | Class (T2) | Model output | Actual fault |
|---|---|---|---|---|---|---|---|---|---|---|---|
| 1 | 32,930 | 2397 | 157 | 0.01 | 0.01 | 0.393 | 0.770 | 0.299 | 0.595 | PD | PD |
| 2 | 9340 | 995 | 60 | 6 | 7 | 0.406 | 0.780 | 0.300 | 0.553 | PD | PD |
| 3 | 78 | 20 | 11 | 13 | 28 | 0.983 | 0.087 | 0.031 | 0.035 | D | D |
| 4 | 6870 | 1028 | 79 | 900 | 5500 | 0.770 | 0.112 | 0.216 | 0.152 | D | D |
| 5 | 10,092 | 5399 | 530 | 6500 | 37,565 | 0.759 | 0.131 | 0.236 | 0.187 | D | D |
| 6 | 8800 | 64,064 | 72,128 | 95,650 | 0.01 | 0.172 | 0.131 | 0.059 | 0.954 | T2 | T2 |
| 7 | 100 | 200 | 110 | 670 | 11 | 0.117 | 0.181 | 0.057 | 0.965 | T2 | T2 |
| 8 | 1270 | 3450 | 520 | 1390 | 8 | 0.266 | 0.124 | 0.435 | 0.563 | T2 | T1 & T2 |
| 9 | 66 | 60 | 2 | 7 | 0.0001 | 0.189 | 0.330 | 0.677 | 0.280 | T1 | T1 & T2 |
| 10 | 2310 | 149 | 20 | 3 | 0.001 | 0.423 | 0.776 | 0.310 | 0.567 | PD | T1 & T2 |

## 6. Conclusion and perspectives

This study introduces a resource-efficient VQC framework for transformer fault diagnosis using DGA data, providing a theoretically rigorous and practically viable alternative to classical ML and existing hybrid quantum–classical models. Unlike recent approaches that inadvertently misapply quantum encoding principles and rely heavily on classical post-processing layers, the proposed VQC leverages a physically consistent hybrid feature map that efficiently embeds four key DGA features into a compact two-qubit quantum state. By integrating full entanglement and an EfficientSU2 ansatz within a shallow-depth circuit, the model maintains strong expressive power while minimizing circuit complexity, aligning directly with the hardware limitations of NISQ devices.

Beyond performance metrics, this work demonstrates that well-engineered quantum feature encodings can achieve superior scalability, data efficiency, and hardware feasibility without sacrificing classification accuracy. The proposed VQC framework provides a transparent and physically realizable pathway toward quantum-enhanced predictive maintenance solutions for power system asset management, marking a significant advancement in bridging quantum machine learning theory with its practical deployment in critical industrial applications.

Despite these strengths, several technical constraints remain. The current design has been validated only on moderate-sized tabular DGA

datasets and limited two-qubit circuits; extending it to larger multi-sensor or temporal datasets will require scalable encodings and more efficient parameter-transfer strategies. Hardware experiments were conducted under simplified noise assumptions, and real-device deployment will demand continuous calibration and hybrid error mitigation. These constraints do not diminish the core findings but rather delineate the realistic boundary of present-day quantum advantage in industrial diagnostics.

In future research, we will explore (i) hierarchical or composite feature-map structures to capture higher-order fault interactions, (ii) quantum–classical feedback calibration for adaptive model tuning under noise, and (iii) hardware-in-the-loop workflows enabling persistent learning in field environments. Collectively, these directions aim to evolve the present prototype into a sustainable, quantum-assisted predictive-maintenance framework, bridging the conceptual gap between laboratory-level quantum learning and real-world power-system reliability.

**CRediT authorship contribution statement**

**Dai Huynh:** Validation, Software, Resources, Formal analysis, Data curation. **Ba Tu Phung:** Writing – original draft, Validation, Software, Resources, Formal analysis, Data curation. **Kim-Anh Nguyen:** Writing – review & editing, Writing – original draft, Visualization, Validation, Supervision, Software, Resources, Project administration, Methodology, Investigation, Funding acquisition, Formal analysis, Data curation, Conceptualization. **Huy Hoang Le:** Writing – review & editing, Writing – original draft, Visualization, Validation, Supervision, Software, Methodology, Investigation, Formal analysis, Conceptualization.

**Declaration of Competing Interest**

The authors declare that they have no known competing financial interests or personal relationships that could have appeared to influence the work reported in this paper.

## Appendix A. From proposed feature map circuit to closed-form kernel (one data re-upload)

Because $R_{YY}(\theta)$ and $R_{ZX}(\theta)$ do not commute, their ordered product injects feature mixing even without explicit $x_i x_j$terms. So the feature map is:

$$|\psi(\vec{x}^{\cdot})\rangle = U_{feature_map}(\vec{x}^{\cdot})|00\rangle = R_{ZX}(x_3^{\cdot})_{0,1} R_{YY}(x_1^{\cdot})_{1,0} R_{ZX}(x_2^{\cdot})_{1,0} R_{YY}(x_0^{\cdot})_{0,1}|00\rangle. \tag{A.1}$$

We use the standard Pauli-rotation identity $R_P(\theta) = e^{-i\frac{\theta}{2}P} = \cos\frac{\theta}{2}I - i\sin\frac{\theta}{2}P, \quad P^2 = I$ and write $c_k = \cos(x_k/2),\ s_k = \sin(x_k/2),$ and$c_k^{'},\ s_k^{'}$ for $z$.

**Step 1 -** apply $R_{YY}(x_0)_{0,1}$to $|00\rangle$:

Since $(Y\otimes Y)|00\rangle = -|11\rangle$,wehaveobtains $R_{YY}(x_0)|00\rangle = c_0|00\rangle + is_0|11\rangle$.

**Step 2** - apply $R_{ZX}(x_2)_{1,0}$ (i.e., $X\otimes Z$):

Using $R_{ZX}(x_2)|00\rangle = c_2|00\rangle - is_2|10\rangle$ and $R_{ZX}(x_2)|11\rangle = c_2|11\rangle + is_2|01\rangle$, we obtain

$|\phi\rangle = c_0c_2|00\rangle - ic_0s_2|10\rangle + is_0c_2|11\rangle - s_0s_2|01\rangle$.

**Step 3** - apply $R_{YY}(x_1)_{1,0}$:

With $\begin{bmatrix} R_{YY}(x_1)|00\rangle = c_1|00\rangle + is_1|11\rangle, & R_{YY}(x_1)|11\rangle = is_1|00\rangle + c_1|11\rangle, \\ R_{YY}(x_1)|01\rangle = c_1|01\rangle - is_1|10\rangle, & R_{YY}(x_1)|10\rangle = -is_1|01\rangle + c_1|10\rangle, \end{bmatrix}$

we define the "thermal" channel $C_y = \cos\frac{x_0+x_1}{2} = c_0c_1 - s_0s_1,\quad S_y = \sin\frac{x_0+x_1}{2} = s_1c_0 + c_1s_0$, and simplify to $|\chi\rangle = C_yc_2|00\rangle - iC_ys_2|10\rangle - S_ys_2|01\rangle + iS_yc_2|11\rangle$.

**Step 4** - apply $R_{ZX}(x_3)_{0,1}$ (i.e., $Z\otimes X$):

Using $\begin{bmatrix} R_{ZX}(x_3)|00\rangle = c_3|00\rangle - is_3|01\rangle, & R_{ZX}(x_3)|01\rangle = c_3|01\rangle - is_3|00\rangle, \\ R_{ZX}(x_3)|10\rangle = c_3|10\rangle + is_3|11\rangle, & R_{ZX}(x_3)|11\rangle = c_3|11\rangle + is_3|10\rangle, \end{bmatrix}$

we obtain the final state

$$\begin{aligned} \psi(x)\rangle &= (c_2c_3C_y + is_2s_3S_y)|00\rangle + (-ic_2s_3C_y - c_3s_2S_y)|01\rangle \\ &+(-ic_3s_2C_y - s_3c_2S_y)|10\rangle + (s_3s_2C_y + ic_3c_2S_y)|11\rangle \end{aligned} \tag{A.2}$$

*Inner product and kernel*

For another input $z$with $C_y^{'}, S_y^{'}, c_2^{'}, s_2^{'}, c_3^{'}, s_3^{'}$, the inner product factors cleanly via half-angle identities:

$$\langle\psi(x)|\psi(z)\rangle = \underbrace{\left(C_yC_y^{'} + S_yS_y^{'}\right)}_{\cos\frac{\Delta_y}{2}}\underbrace{(c_2c_2^{'} + s_2s_2^{'})}_{\cos\frac{\Delta_2}{2}}\underbrace{(c_3c_3^{'} + s_3s_3^{'})}_{\cos\frac{\Delta_3}{2}} + i\underbrace{\left(C_yS_y^{'} - S_yC_y^{'}\right)}_{\sin\frac{\Delta_y}{2}}\underbrace{(c_2s_2^{'} - c_2^{'}s_2)}_{\sin\frac{\Delta_2}{2}}\underbrace{(c_3s_3^{'} - c_3^{'}s_3)}_{\sin\frac{\Delta_3}{2}}, \tag{A.3}$$

where $\Delta_y = (x_0 + x_1) - (z_0 + z_1);\ \Delta_2 = x_2 - z_2;\ \Delta_3 = x_3 - z_3$. Therefore the fidelity kernel is

$$K(x,z) = |\langle\psi(x)|\psi(z)\rangle|^2 = \cos^2\left(\frac{\Delta_y}{2}\right)\cos^2\left(\frac{\Delta_2}{2}\right)\cos^2\left(\frac{\Delta_3}{2}\right) + \sin^2\left(\frac{\Delta_y}{2}\right)\sin^2\left(\frac{\Delta_2}{2}\right)\sin^2\left(\frac{\Delta_3}{2}\right). \tag{A.4}$$

The equivalent form $K(x,z)$ shows explicit pairwise interactions among the three channels $(\Delta_y, \Delta_2, \Delta_3)$, which do not arise with separable Angle Encoding / Amplitude Encoding and differ from commuting ZZ maps. The proposed feature map (Eq. A.5) and Kernel (Eq. A.6) are shown below:

$$|\psi(\vec{x}^{\cdot})\rangle = U_{feature_map}(\vec{x}^{\cdot})|00\rangle = R_{ZX}(x_3^{\cdot})_{0,1} R_{YY}(x_1^{\cdot})_{1,0} R_{ZX}(x_2^{\cdot})_{1,0} R_{YY}(x_0^{\cdot})_{0,1}|00\rangle, \tag{A.5}$$

$$K(x,z) = |\langle\psi(x)|\psi(z)\rangle|^2 = \cos^2\left(\frac{\Delta_y}{2}\right)\cos^2\left(\frac{\Delta_2}{2}\right)\cos^2\left(\frac{\Delta_3}{2}\right) + \sin^2\left(\frac{\Delta_y}{2}\right)\sin^2\left(\frac{\Delta_2}{2}\right)\sin^2\left(\frac{\Delta_3}{2}\right). \tag{A.6}$$

That said, we can calculate frame potential $F_2 = E_{x,z}[K(x,z)^2]$ and $F_2^{Haar} = \frac{2}{N(N+1)}$, with $N = 4 \rightarrow$ $F_2^{Haar} = 0.1$ to estimated expressivity using Monte-Carlo method $\widehat{F}_2 = \frac{1}{M}\sum_{m=1}^{M} K(x^{(m)}, z^{(m)})^2$, $x^{(m)}$, $z^{(m)} \sim \mu([-\pi,\pi]^4)$. Lower $F_2$ (closer to $F_2^{Haar}$) indicates a richer state ensemble [46], suggesting that the encoding circuit can encompass a substantial portion of the Hilbert space. However, this does not imply that minimal expressiveness necessarily results in superior feature encoding; rather, it signifies that the feature map can represent a broader range within the Hilbert space. Fig. A.1 indicates the expressivity between feature maps, encoder-only with re-upload depth $L$, error bars are 95% CIs over 5000 pairs. As practicing, we only use the first layer $L = 1$for encoding features, which our proposed feature map achieve the best expressivity via lowest $F_2$ and closest to $F_2^{Haar}$.

As shown in Fig. A.1 and summarized in Table A.1, the $ZX/YY$-hybrid feature map offers the best balance between expressivity and trainability for our task. Its KTA = 0.520 indicates that the induced kernel aligns well with the label structure, while its frame potential $F_2 = 0.108 \pm 0.003$ at $L = 1$ sits close to the Haar limit (0.10), meaning the encoding explores a sufficiently rich portion of Hilbert space without pushing depth. At the same time, the gradient variance (0.244) is large enough to prevent vanishing-gradient issues yet moderate enough to keep optimization stable, and the effective dimension $d_{eff}(\Delta) = \mathrm{Tr}\left[F(F+\Delta I)^{-1}\right]$ matches the most expressive baselines ($d_{eff} = 13$).

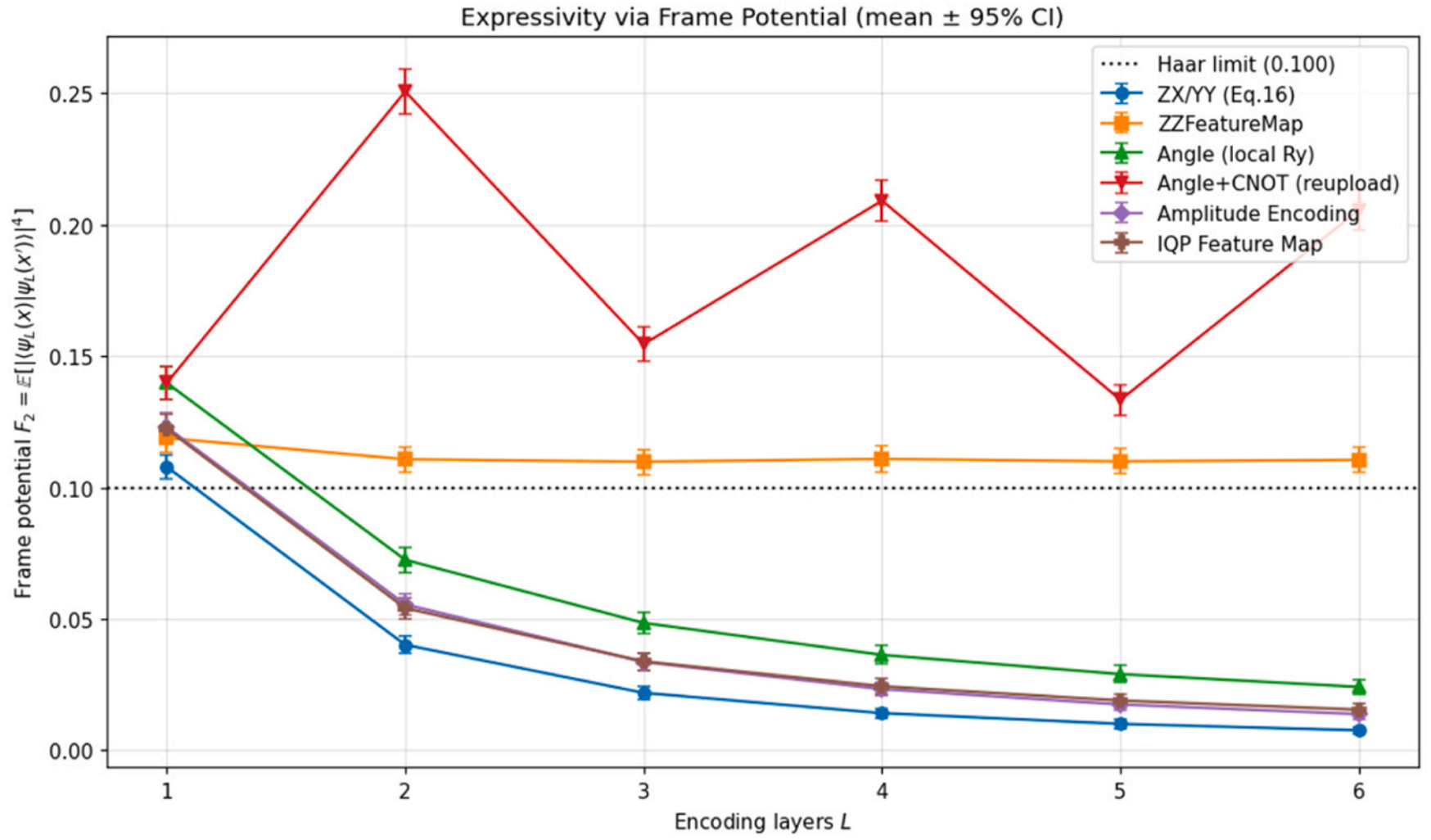

**Fig. A.1.** Expressivity between feature maps via frame potential (mean ± 95% CI)

Specifically, the "entanglement (bits)" metric in Table A.1 is defined as the average von Neumann entropy of the reduced single-qubit density matrix: $S(\rho_A) = -\mathrm{Tr}(\rho_A \log_2 \rho_A)$, where $\rho_A = \mathrm{Tr}_B(|\psi\rangle\langle\psi|)$is obtained by tracing out qubit B from the full 2-qubit state $|\psi\rangle$. The average is taken over all test-set inputs passed through the trained feature map. This metric ranges from 0 (product state, no entanglement) to 1 bit (maximally entangled Bell state). The value of 0.59 bits for the proposed ZX/YY-hybrid indicates moderate entanglement, which is sufficient to generate the pairwise interaction terms identified in the kernel analysis, while remaining in a regime that is compatible with current NISQ noise levels (avoiding the highly noise-sensitive regimes of high-entanglement maps like IQP or ZZ).

**Table A.1**
Characterization of feature maps by separability and trainability

| Feature Map | KTA | Expressivity ($L = 1$) | Entanglement (bits) | Gradient variance | Effective dimension | Number of features encoded using $n$ qubits |
|---|---|---|---|---|---|---|
| ZX/YY-hybrid | 0.520 | 0.108 ± 0.003 | 0.59 | 0.244 | 13 | $2^n$ |
| Angle (local Ry) | 0.608 | 0.137 ± 0.005 | 0.00 | 0.501 | 9 | $n$ |
| Angle + CNOT | 0.203 | 0.140 ± 0.006 | 0.31 | 0.379 | 9 | $n$ |
| Amplitude Encoding | 0.508 | 0.120 ± 0.003 | 0.26 | 0.412 | 10 | $2^n$ |
| IQP Feature Map | 0.407 | 0.121 ± 0.003 | 0.77 | 0.113 | 13 | $n$ |
| ZZFeatureMap | 0.366 | 0.117 ± 0.002 | 0.80 | 0.134 | 13 | $n$ |

Mathematically, the closed-form kernel introduces cross-terms in the input differences $\Delta$ that act like higher-order interactions in $a$ classical feature space, thereby improving separability, and it does so with an encoding capacity that scales as $2^n$features on $n$ qubits. Taken together, these properties explain why we fix $ZX/YY$ with $L = 1$ in both simulation and IBM hardware runs: it is expressive enough to separate classes, shallow and robust enough for real devices, and directly compatible with the native gate set after transpilation.

We also consider complexity and gate-efficiency trade-offs:

- Gate / depth complexity: One data upload uses 4 two-qubit rotations $YY, ZX, YY, XZ$. Each $\exp(-i\theta P \otimes Q/2)$ compiles to ~2 CNOTs + 1–2 1q rotations (backend-dependent), so the encoder costs ~8 CNOTs total. Depth is constant in input size (4 DGA features → one layer) and comparable to 4-layers EfficientSU2 block.
- Head-to-head gate trade-offs: Table A.2 reports {$G_2$, 1q-rotates, depth}) for our $ZX/YY$, Angle ($R_Y(\theta)$), Angle+CNOT re-upload, ZZFeatureMap, IQP, all with the same mixer depth (4 ×EfficientSU2). This shows that although $ZX/YY$ pays ~8 CNOTs at upload, it induces non-commuting pairwise channels at encoding, so a single SU2 layer suffices-yielding lower total $G_2$at equal accuracy than separable/commuting uploads that require deeper mixing, which $G_2$ is the number of CNOTs used.

**Table A.2**
Feature map's gate-efficiency trade-offs

| Encoder (one upload) | 2q / upload | 1q / upload | Kernel character | End-to-end 2q | Trainable parameters | $F_2$ |
|---|---|---|---|---|---|---|
| ZX–YY (proposed) | 8 | 8 | Non-commuting, entangling (pairwise 20channels) | 8 + 4 = 12 | 20 | 0.101 ± 0.003 |
| Angle | 0 | 2 | Separable (no entanglement) | 0 + 4 = 4 | 20 | 0.124 ± 0.016 |
| Angle + CNOT re-upload | 2 | 2 | Weak interaction (via CX) | 2 + 4 = 6 | 20 | 0.148 ± 0.051 |
| ZZFeatureMap | 2 | 2 | Commuting Z-phases | 2 + 4 = 6 | 20 | 0.116 ± 0.042 |
| IQP (H–diag–H) | 2 | 6 | Diagonal-phase kernel | 2 + 4 = 6 | 20 | 0.106 ± 0.035 |

## Data availability

The data that support the findings of this study are openly available in the published literature and were obtained from [47] and [48]. No new datasets were generated or analyzed during the current study. The implementation code and trained model parameters supporting this study are openly available, without access restrictions, at https://github.com/HuyHoangLe0201/VQC-DGA.